\documentclass[12pt]{article}
\usepackage[colorlinks=true,citecolor=blue,linkcolor=blue]{hyperref}
\usepackage{multirow}
\usepackage[left=2cm,right=2cm,top=2cm,bottom=2cm]{geometry}
\usepackage{cite}
\usepackage{psfrag}
\usepackage{graphicx}
\usepackage{graphics}
\usepackage{epsfig}
\usepackage{amsmath,amsfonts,amssymb}
\usepackage{epstopdf}
\usepackage{pstcol,pst-fill,pst-grad}
\usepackage{euscript}
\usepackage{pstricks}
\usepackage{wrapfig}
\usepackage{subfig}
\usepackage{slashed}
\usepackage[toc,page]{appendix}
\usepackage{here}

\date{}
\begin{document}
	\title{\vspace{-3cm}
		\hfill\parbox{4cm}{\normalsize \emph{}}\\
		\vspace{1cm}
		{ Relativistic electron-impact ionization of hydrogen atom from its metastable 2S-state in the symmetric/asymmetric coplanar geometries}}
	\vspace{2cm}
	
	\author{ M Jakha$^1$, S Mouslih$^2$, S Taj$^1$, and B Manaut$^{1,}$\thanks{Corresponding author, E-mail: b.manaut@usms.ma} \\
		{\it {\small$^1$ USMS, Facult\'e Polydisciplinaire, \'Equipe de  Recherche en Physique Th\'eorique et Mat\'eriaux }}\\
			{\it {\small (ERPTM), B\'eni Mellal, 23000, Morocco.}}\\
	{\it {\small$^2$ USMS, Facult\'e des Sciences et Techniques,
Laboratoire de Physique des Mat\'eriaux (LPM),}}\\
	{\it {\small B\'eni Mellal, 23000, Morocco.}}		
	}
	\maketitle \setcounter{page}{1}
\begin{abstract}
We analytically compute, in the first Born approximation for symmetric and asymmetric coplanar geometries, the triple differential cross sections for electron-impact ionization of hydrogen atom in the metastable 2S-state at both low and high energies. The process is investigated by using the relativistic Dirac-formalism and it is also shown that the nonrelativistic limit is accurately reproduced when using low incident kinetic energies.  At high energies, relativistic and spin effects significantly affect the triple differential cross sections. Our analytical approach which seems exact is compared to some other results in the nonrelativistic regime for asymmetric coplanar geometry. For this particular process and in the absence of any experimental data and theoretical models at high energies, we are not in a position to validate our model. We hope that the present study will provide significant contribution to future experiments.
\vspace{.04cm}\\

\vspace{.04cm}\\

\end{abstract}
\section{Introduction}
In recent years, much attention has been paid to the experimental and theoretical aspects of collision processes involving metastable atoms. This is principally due to the fact that atoms in metastable states own some  properties such as long lifetimes,  ability to transmit large amounts of energy, low excitation and ionization potentials, resulting in very large cross-sections. From the collision point of view, the ionization of metastable atoms is very important to understand the mechanisms that occur in astrophysical and fusion plasmas, in partially ionized systems, and also play a major role in the gas discharge phenomenon. Apart from all of these, metastable states of atoms are nowadays gaining increasing importance in many areas of research, for example, cold atomic physics, in particular Bose-Einstein condensation, nanolithography and also famously in laser physics \cite{ghoshdeb}. Electron-impact ionization is the removal of one or more electrons from the target resulting from the collision between it and an electron. We can distinguish different types of ionization; single ionization, called (e, 2e) process, which occurs when the resulting ion leaves the collision region with a single positive charge, multiple ionization where several electrons in the electronic cortege are ejected and the ion can have multiple positive charges. In this work, however, we will only deal with the case of single ionization of the hydrogen atom from its metastable 2S-state, when it is bombarded by an electron of energy $E_{i}$ greater than the ionization potential. In the collision zone, two electrons emerge with energies $E_{f}$ and $E_{B}$. Even though these two electrons cannot be distinguished, it is convenient to call the faster electron "scattered electron" and the slower one "ejected electron". Electron-impact ionization of atomic, ionic or molecular systems is one of the important processes of collisional physics, in particular for the study of the structure of matter. It also finds its application in various fields such as astrophysics and plasma physics.
Especially, the electron-impact single ionization has proved to be a powerful tool for studying the structure of atoms and their dynamics. Ionization of hydrogen atoms by electron impact is the fundamental and simplest ionization process. The hydrogen atom is an ideal target due to its analytically known wave functions, although it is a particularly difficult target for experimentalists. At present, there are many theoretical models to compute the cross sections of hydrogen-atoms ionization in both the ground and metastable states at various incident kinetic energies and under different kinematic conditions. Unfortunately, ionization from metastable states has not been investigated to the same extent, especially in the relativistic regime, as ionization from the ground state; and this is mainly due to the lack of any experimental studies on this type of ionization. The investigation of the ionization from metastable states of hydrogen atoms by charged particles is now equally interesting and experimental results will soon be available in this field. In particular, the fully triple differential cross sections (TDCS) for the (e, 2e) process have been extensively studied for the ground state hydrogen atom both theoretically \cite{taj,attaourti,nakel,brauner1,brauner2,berkader,kover1,kover2} and experimentally \cite{dorr,dorn,ren}, while for the ionization from the metastable state no such measurement of TDCS is yet available in the literature, although the absolute total cross sections have been measured much earlier \cite{dixon1975,defrance1981}. However, on the theoretical side, quite a few calculations have been performed on the TDCS of the metastable (2S) hydrogen atom using electron impact \cite{hafid,vucic,BBK,dhar,das,biswas,ray} and significant differences were observed in the TDCS structures when compared to the cross section of ground-state ionization. All these theoretical calculations available in the literature to date have been done within the framework of the asymmetric geometry and at low energies. To the best of our knowledge, there is no study available to the ionization of the hydrogen atom from its metastable 2S-state using relativistic formalism at high energies. This work addresses, for the first time, a theoretical study and an analytical calculation of the ionization of the hydrogen atom from its metastable 2S-state at high energies in both symmetric and asymmetric coplanar geometries taking into account the effects of spin and relativity. In the asymmetric coplanar geometry, we present a theoretical semirelativistic Coulomb Born approximation (SRCBA) for the description of the ionization of hydrogen atom by electron impact in the first Born approximation. In this approximation, the incident and scattered electrons are described by Dirac plane relativistic wave functions while the ejected electron is described by a Sommerfeld-Maue semirelativistic Coulomb wave function and the hydrogen atom, in its metastable state, is described by Darwin's semirelativistic wave function. The TDCS obtained in SRCBA will be compared with the corresponding one in the nonrelativistic Coulomb Born approximation (NRCBA). In the symmetric coplanar geometry, we present the relativistic formalism of the (e, 2e) reaction in the relativistic plane wave Born approximation (RPWBA), where the incident, scattered and ejected electrons are described by relativistic plane waves, and the hydrogen atom in its metastable 2S-state is described by the relativistic exact function, and it will be compared, in the nonrelativistic
domain, with the nonrelativistic plane wave Born
approximation (NRPWBA). We confirm here that our work is in fact an extension of a work published in $2005$ \cite{taj} by two of our co-authors through which, for the first time, the SRCBA model was applied to the ground state of the hydrogen atom and proved its validity compared to experimental results and results of other theoretical approaches. The same is true for the RPWBA model. A detailed account of the TDCS in the ground state of the hydrogen atom was also presented, based on a relativistic formalism, in the paper \cite{attaourti}, and its validity was proven by its comparison with the NRPWBA model in the non-relativistic domain due to the absence of results at high energies. If these two models gave good results in the ground state of the hydrogen atom, then the results obtained in the metastable 2S-state are also expected to be good and acceptable. 
The only difference between our work and the previous ones \cite{attaourti,taj} is in the wave function that describes the hydrogen atom in all computed models. This one difference is capable enough to completely change the calculation and give rise to new analytically complex integrals. We would like to point out here that our goal in this research is not exactly to study ionization process in the non-relativistic regime, but rather to study it at high energies by applying the relativistic Dirac formalism. Thus, the NRCBA model that we have calculated in this paper is only a way to compare the TDCS obtained in the SRCBA model. There are many complex approaches that are applied when studying ionization process in the non-relativistic domain, we mention here as example, the R-matrix (RM) theory \cite{rmatrix}, the Convergent Close-Coupling (CCC) method \cite{ccc}, and the distorted-wave Born approximation (DWBA) \cite{dwba}. Since our goal is to study this reaction at high energies, we decided to be satisfied only, in the non-relativistic domain, with the application of the NRCBA model due to the relative simplicity of its Coulomb wave function. We remind here that, in all our calculations in the various models including the NRCBA model, we did not take into account the residual ion H$^{+}$, nor the exchange effects (since we work at high energies in which the exchange effects are negligible) or the various interactions that may occur between the two final electrons. We have found that the relativistic and spin effects, become more and more important by increasing the energy of the incident electron. All the appropriate numerical tests to verify the validity of the analytical results we found were performed with a very good degree of accuracy. The paper is constructed as follows. In section 2, we deliver the different theoretical models in the asymmetric and symmetric coplanar geometries and give, for each model, a detailed account of the techniques which we have used to evaluate the TDCS. In section 3, we discuss the numerical results we have obtained in each geometry. Finally, section 4 is devoted to the conclusions. Atomic units $\hbar=m=e=1$ are used throughout this work.
\section{Theoretical models}
Let us consider a collision between a hydrogen atom in its metastable 2S-state and an incident electron moving along the $z$-axis. As a result of this collision, the hydrogen atom becomes ionized and the projectile electron changes its four-momentum from $p_{i}$ to $p_{f}$. In the final state, two electrons (scattered and ejected) emerge with four-momenta $p_{f}$ and $p_{B}$. This reaction can be described, symbolically, as follows:
\begin{equation}\label{process}
\text{e}^{-}(p_{i})+\text{H(2S)}\longrightarrow \text{H}^{+}+\text{e}^{-}(p_{B})+\text{e}^{-}(p_{f}).
\end{equation}
All ionization reactions are studied within two geometric frameworks. The first is called asymmetric geometry and the second is symmetric geometry, and each of them may be coplanar or noncoplanar. In asymmetric geometries, a fast electron of energy $E_{i}$ is incident on the target atom, and a fast scattered electron is detected in coincidence with a slow ejected electron. This kind of experiment was first performed by Ehrhardt \textit{et al.} \cite{ehrhardt1969}. Symmetric geometries, which are defined by the requirement that the two outgoing electrons are detected with the same energies and equal scattering and ejection angles  (i.e.  $E_{f}\simeq E_{B}$ and $\theta_{f}\simeq\theta_{B}$ ), were introduced by Amaldi \textit{et al.} \cite{amaldi1969}. In coplanar geometry, the three momenta $\textbf{p}_{i}$, $\textbf{p}_{f}$ and $\textbf{p}_{B}$ are in the same plane, whereas in noncoplanar geometry the momentum $\textbf{p}_{B}$ is out of the ($\textbf{p}_{i}$, $\textbf{p}_{f}$) reference plane.
\begin{figure}[hbtp]
	\centering
	\includegraphics[scale=0.4]{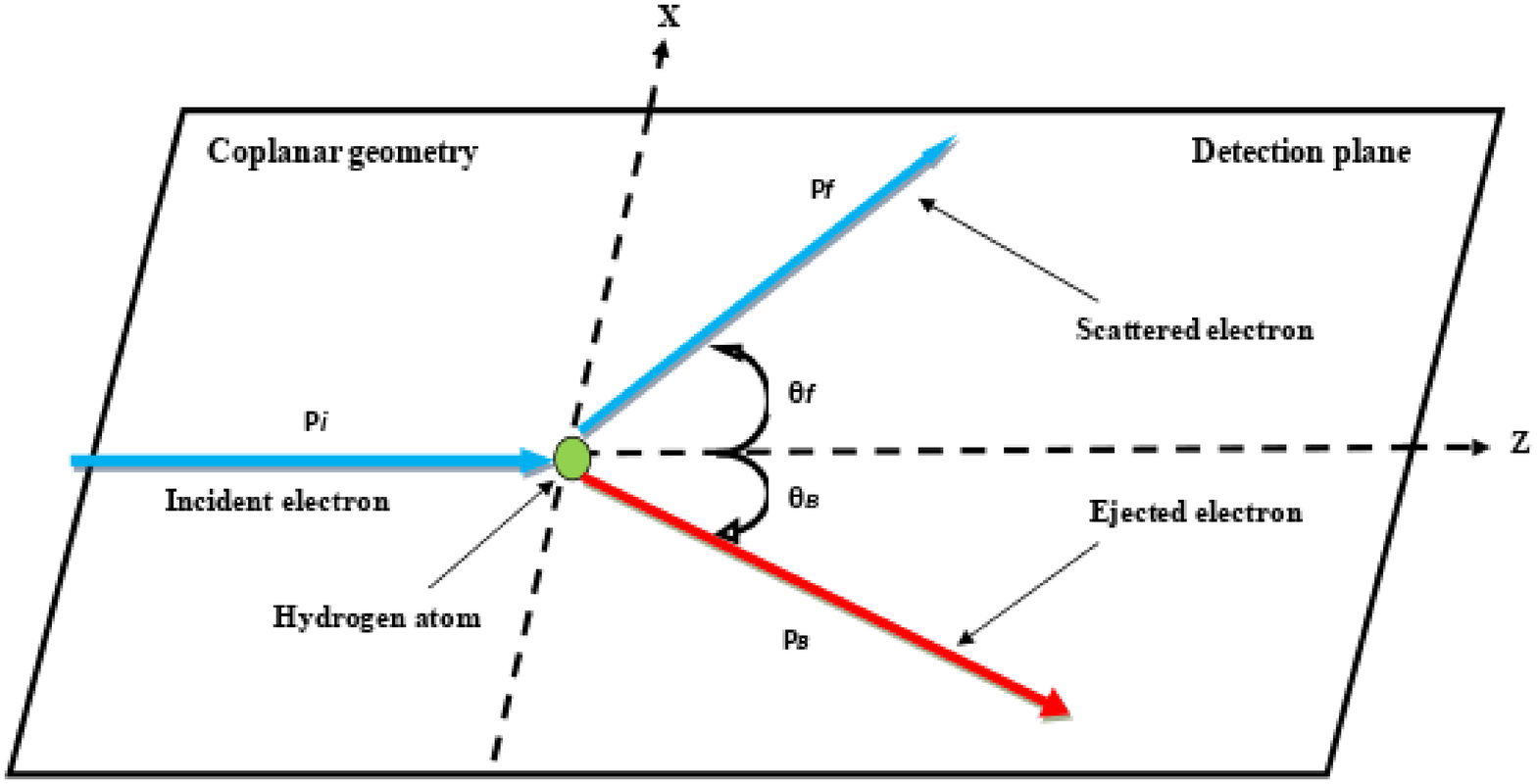}
	\caption{Schematic diagram of the coplanar geometry chosen for our theoretical study. $\theta_{f}$ and $\theta_{B}$ are, respectively, the angles of the scattered and ejected electrons with respect to the incident beam direction $\textbf{p}_{i}$. For coplanar symmetric geometries, $\theta_{f}\simeq\theta_{B}$ .}
\end{figure}\\
During this work, we will study the process (\ref{process}) under two different geometries. We will start first with asymmetric coplanar geometry and then secondly with symmetric coplanar geometry. The detailed calculation of each TDCS in each geometry will be presented.
\subsection{Asymmetric coplanar  geometry}
We remember that in the case of the Ehrhardt coplanar asymmetric geometry, a fast electron of kinetic energy $T_{i}$ is incident on the hydrogen target, and a fast scattered electron of kinetic energy $T_{f}$ is detected in coincidence with a slow ejected electron of kinetic energy $T_{B}$. Additionally, the three momenta $\textbf{p}_{i}$, $\textbf{p}_{f}$, and $\textbf{p}_{B}$ are in the same plane and the scattering angle $\theta_{f}$ of the scattered electron is fixed and small, while the angle $\theta_{B}$ of the ejected electron is varied. In this geometry, we calculate step by step the exact analytical expression of the semirelativistic spin-unpolarized TDCS in the SRCBA approximation corresponding to the electron-impact ionization of atomic hydrogen in its metastable 2S-state.
\subsubsection{The S-matrix element}
We begin with the first Born ionization S-matrix element for the process (\ref{process}) in the direct channel in which the exchange effects are neglected. It can be written as \cite{book1}
\begin{equation}\label{smatrix}
\begin{split}
S_{fi}&=-\frac{i}{c}\int_{-\infty}^{+\infty}dx^{0}\langle\psi_{p_{f}}(x_{1})\phi_{f}(x_{2})|V_{d}|\psi_{p_{i}}(x_{1})\phi_{i}(x_{2})\rangle,\\
&=-i\int_{-\infty}^{+\infty}dt \int d\textbf{r}_{1}\psi_{p_{f}}^{\dag}(t,\textbf{r}_{1})\psi_{p_{i}}(t,\textbf{r}_{1})\langle\phi_{f}(x_{2})|V_{d}|\phi_{i}(x_{2})\rangle.
\end{split}
\end{equation}
Here, the potential $V_{d}=1/r_{12}-1/r_{1}$ presents the direct interaction between the incident electron and the hydrogen atom, where $r_{12}=|\textbf{r}_{1}-\textbf{r}_{2}|$ and $r_{1}=|\textbf{r}_{1}|$. The nucleus of the target atom, which is assumed to be infinitely massive, is chosen to be the origin of the coordinate system. The coordinates of the incident and atomic electrons are labeled by $\textbf{r}_{1}$ and $\textbf{r}_{2}$, respectively. $\psi_{p_{i}}$ and $\psi_{p_{f}}$ are the wave functions describing, respectively, the incident and scattered electrons given by a free Dirac solution normalized to the volume $V$
\begin{equation}
\begin{split}
\psi_{p_{i}}(x_{1})&=\frac{u(p_{i},s_{i})}{\sqrt{2E_{i}V}}e^{-ip_{i}.x_{1}},\\
\psi_{p_{f}}(x_{1})&=\frac{u(p_{f},s_{f})}{\sqrt{2E_{f}V}}e^{-ip_{f}.x_{1}},
\end{split}
\end{equation}
where $E_{i}$ and $E_{f}$ are, respectively, the total energies of the incident and scattered electrons.
$\phi_{i}(x_{2})=\phi_{i}(t,\textbf{r}_{2})$ is the semirelativistic Darwin wave function of atomic hydrogen in its metastable 2S-state, which is accurate to the order $Z/c$ in the relativistic corrections. It is given by
\begin{equation}
\phi_{i}(t,\textbf{r}_{2})=\exp[-i\mathcal{E}_{b}(2S) t]\varphi^{(\pm)}_{2S}(\textbf{r}_{2}),
\end{equation}
where $\mathcal{E}_{b}(2S)$ is the binding energy of the metastable 2S-state of atomic hydrogen given by
\begin{equation}\label{binding}
\mathcal{E}_{b}(2S)=\frac{c^{2}}{\sqrt{2}}\sqrt{1+\sqrt{1-\alpha^{2}}}-c^{2},
\end{equation}
where $\alpha=1/c$ is the fine structure constant. For spin up, $\varphi^{(+)}_{2S}(\textbf{r}_{2})$ is expressed by
\begin{equation}\label{Darwin}
\varphi^{(+)}_{2S}(\textbf{r}_{2})=N_{D_{2}}\begin{pmatrix} 2-r_{2} \\ 0 \\ \frac{i(4-r_{2})}{4c}\cos(\theta) \\ \frac{i(4-r_{2})}{4c}\sin(\theta)e^{i\phi}\end{pmatrix}\frac{1}{4\sqrt{2\pi}}e^{-r_{2}/2},
\end{equation}
where $\theta$ and $\phi$ are the spherical coordinates of $\textbf{r}_{2}$ and
\begin{equation}
N_{D_{2}}=\frac{4c}{\sqrt{32c^{2}+10}}
\end{equation}
is the normalization constant. The wave function $\phi_{f}(x_{2})=\phi_{f}(t,\textbf{r}_{2})$ in Eq.~(\ref{smatrix}) is the Sommerfeld-Maue wave function for continuum states \cite{book1}, also accurate to the order $Z/c$ in the relativistic corrections. We have
\begin{equation}
\phi_{f}(t,\textbf{r}_{2})=e^{-iE_{B}t} \psi_{p_{B}}^{(-)}(\textbf{r}_{2}),
\end{equation}
where $\psi_{p_{B}}^{(-)}(\textbf{r}_{2})$ is given, in its final compact form normalized to the volume V, by
\begin{equation}\label{maue}
\begin{split}
\psi_{p_{B}}^{(-)}(\textbf{r}_{2})&=e^{\pi \eta_{B}/2} \Gamma(1+i\eta_{B})e^{i\textbf{p}_{B}.\textbf{r}_{2}}\bigg \lbrace {}_1F_{1}(-i\eta_{B},1,-i(p_{B}r_{2}+\textbf{p}_{B}.\textbf{r}_{2}))+\frac{i}{2cp_{B}}(\alpha.\textbf{p}_{B}+p_{B}\alpha.\hat{\textbf{r}}_{2})\\&{}_1F_{1}(-i\eta_{B}+1,2,-i(p_{B}r_{2}+\textbf{p}_{B}.\textbf{r}_{2}))\bigg\rbrace\frac{u(p_{B},s_{B})}{\sqrt{2E_{B}V}}.
\end{split}
\end{equation}
$\eta_{B}$ is the Sommerfeld parameter given by
\begin{equation}
\eta_{B}=\frac{E_{B}}{c^{2}p_{B}},
\end{equation}
where $E_{B}$ is the total energy of the ejected electron and $p_{B}=|\textbf{p}_{B}|$ is the norm of the ejected electron momentum. In Eq.~(\ref{maue}), the operator $[\alpha.\textbf{p}_{B}]$ acts on the free spinor $u(p_{B},s_{B})$ and the operator $[\alpha.\hat{\textbf{r}}_{2}]$ acts on the spinor part of the Darwin wave function.\\
The integral over the time coordinate in Eq.~(\ref{smatrix}) can be separated yielding
\begin{equation}
\int dt \exp[i(E_{f}+E_{B}-E_{i}-\mathcal{E}_{b}(2S))t]=2\pi\delta(E_{f}+E_{B}-E_{i}-\mathcal{E}_{b}(2S)),
\end{equation}
while the integration over $d\textbf{r}_{1}$ can be performed by using the well-known following Bethe integral
\begin{equation}
\int d\textbf{r}_{1} e^{i(\textbf{p}_{i}-\textbf{p}_{f}).\textbf{r}_{1}}\bigg(\frac{1}{r_{12}}-\frac{1}{r_{1}}\bigg)=\frac{4\pi}{\Delta^{2}}(e^{i\mathbf{\Delta}.\textbf{r}_{2}}-1),
\end{equation}
where the quantity $\mathbf{\Delta}=\textbf{p}_{i}-\textbf{p}_{f}$ is the momentum transfer.\\
The direct S-matrix element in Eq.~(\ref{smatrix}) becomes
\begin{equation}\label{smatrixp}
\begin{split}
S_{fi}&=-i\int d\textbf{r}_{2} \frac{\bar{u}(p_{f},s_{f})}{\sqrt{2E_{f}V}}\gamma^{0}\frac{u(p_{i},s_{i})}{\sqrt{2E_{i}V}}\bigg \lbrace {}_1F_{1}(i\eta_{B},1,i(p_{B}r_{2}+\textbf{p}_{B}.\textbf{r}_{2}))-\frac{i}{2cp_{B}}(\alpha.\textbf{p}_{B}\\&+p_{B}\alpha.\hat{\textbf{r}}_{2}){}_1F_{1}(i\eta_{B}+1,2,i(p_{B}r_{2}+\textbf{p}_{B}.\textbf{r}_{2}))\bigg\rbrace \frac{\bar{u}(p_{B},s_{B})\gamma^{0}}{\sqrt{2E_{B}V}}\varphi^{(+)}_{2S}(\textbf{r}_{2})e^{-i\textbf{p}_{B}.\textbf{r}_{2}}(e^{i\mathbf{\Delta}.\textbf{r}_{2}}-1)\\&\times \frac{8\pi^{2}}{\Delta^{2}}\delta(E_{f}+E_{B}-E_{i}-\mathcal{E}_{b}(2S)) e^{\pi \eta_{B}/2} \Gamma(1-i\eta_{B}).
\end{split}
\end{equation}
This S-matrix element contains two terms $S_{fi}^{(1)}$, $S_{fi}^{(2)}$. The first one is given by
\begin{equation}
\begin{split}
S_{fi}^{(1)}&=-i\int d\textbf{r}_{2} \frac{\bar{u}(p_{f},s_{f})}{\sqrt{2E_{f}V}}\gamma^{0}\frac{u(p_{i},s_{i})}{\sqrt{2E_{i}V}}\frac{\bar{u}(p_{B},s_{B})\gamma^{0}}{\sqrt{2E_{B}V}}\big \lbrace {}_1F_{1}(i\eta_{B},1,i(p_{B}r_{2}+\textbf{p}_{B}.\textbf{r}_{2}))\big\rbrace \varphi^{(+)}_{2S}(\textbf{r}_{2})\\&\times e^{-i\textbf{p}_{B}.\textbf{r}_{2}}(e^{i\mathbf{\Delta}.\textbf{r}_{2}}-1) \frac{8\pi^{2}}{\Delta^{2}}\delta(E_{f}+E_{B}-E_{i}-\mathcal{E}_{b}(2S)) e^{\pi \eta_{B}/2} \Gamma(1-i\eta_{B}).
\end{split}
\end{equation}
This first term can be reformulated in the following form
\begin{equation}
\begin{split}
S_{fi}^{(1)}&=-i[H_{1}(\textbf{q}=\mathbf{\Delta}-\textbf{p}_{B})-H_{1}(\textbf{q}=-\textbf{p}_{B})] \frac{\bar{u}(p_{f},s_{f})}{\sqrt{2E_{f}V}}\gamma^{0}\frac{u(p_{i},s_{i})}{\sqrt{2E_{i}V}}\frac{\bar{u}(p_{B},s_{B})\gamma^{0}}{\sqrt{2E_{B}V}}\\&\times\frac{8\pi^{2}}{\Delta^{2}}\delta(E_{f}+E_{B}-E_{i}-\mathcal{E}_{b}(2S)) e^{\pi \eta_{B}/2} \Gamma(1-i\eta_{B}),
\end{split}
\end{equation}
where $H_{1}(\textbf{q})$ is given by
\begin{equation}
H_{1}(\textbf{q})=\int d\textbf{r}_{2} e^{i\textbf{q}.\textbf{r}_{2}}{}_1F_{1}(i\eta_{B},1,i(p_{B}r_{2}+\textbf{p}_{B}.\textbf{r}_{2}))\varphi^{(+)}_{2S}(\textbf{r}_{2}).
\end{equation}
According to the expression of $\varphi^{(+)}_{2S}(\textbf{r}_{2})$ given in Eq.~(\ref{Darwin}), $H_{1}(\textbf{q})$ can be written as
\begin{equation}\label{h1}
H_{1}(\textbf{q})=\frac{N_{D_{2}}}{4\sqrt{2\pi}}(I_{1},0,I_{2},I_{3})^{T},
\end{equation}
and one has to evaluate
\begin{equation}
I_{1}=\int d\textbf{r}_{2} (2-r_{2})e^{-r_{2}/2}e^{i\textbf{q}.\textbf{r}_{2}}{}_1F_{1}(i\eta_{B},1,i(p_{B}r_{2}+\textbf{p}_{B}.\textbf{r}_{2})).
\end{equation}
In this integral, we are confronted with the task of evaluating two types of integrals, one of which is
\begin{equation}
I'_{1}=2\int d\textbf{r}_{2}e^{-r_{2}/2}e^{i\textbf{q}.\textbf{r}_{2}}{}_1F_{1}(i\eta_{B},1,i(p_{B}r_{2}+\textbf{p}_{B}.\textbf{r}_{2})),
\end{equation}
and the other one is
\begin{equation}
I''_{1}=\int d\textbf{r}_{2}r_{2}e^{-r_{2}/2}e^{i\textbf{q}.\textbf{r}_{2}}{}_1F_{1}(i\eta_{B},1,i(p_{B}r_{2}+\textbf{p}_{B}.\textbf{r}_{2})).
\end{equation}
In order to evaluate the two integrals $I'_{1}$ and $I''_{1}$, we take recourse to the well-known integral \cite{integral}
\begin{equation}\label{IntegralI}
\begin{split}
I(\lambda)&=\int d\textbf{r} e^{i\textbf{q}.\textbf{r}}\frac{e^{-\lambda r}}{r}{}_1F_{1}(i\eta_{B},1,i(p_{B}r_{2}+\textbf{p}_{B}.\textbf{r}_{2})),\\
&=\frac{4\pi}{q^{2}+\lambda^{2}}\exp\bigg[i\eta_{B}\ln\bigg(\frac{q^{2}+\lambda^{2}}{q^{2}+\lambda^{2}+2\textbf{q}.\textbf{p}_{B}-2i\lambda p_{B}}\bigg) \bigg],
\end{split}
\end{equation}
where $\lambda$ is a real variable.
Looking at the expressions of the integrals $I'_{1}$ and $I''_{1}$ above, it becomes clear that they are, respectively, the first and second derivatives of the integral $I(\lambda)$. This yields
\begin{equation}
I'_{1}=2\bigg(-\frac{\partial I(\lambda)}{\partial \lambda}\bigg)\bigg|_{\lambda=1/2}.
\end{equation}
\begin{equation}
I''_{1}=\bigg(\frac{\partial^{2} I(\lambda)}{\partial \lambda^{2}}\bigg)\bigg|_{\lambda=1/2}.
\end{equation}
The other integrals $I_{2}$ and $I_{3}$ in (\ref{h1}) can be obtained by noting that
\begin{equation}
\cos(\theta)e^{i\textbf{q}.\textbf{r}_{2}}=-\frac{i}{r_{2}}\frac{\partial}{\partial q_{z}}e^{i\textbf{q}.\textbf{r}_{2}},
\end{equation}
and
\begin{equation}
\sin(\theta)e^{i\phi}e^{i\textbf{q}.\textbf{r}_{2}}=-\frac{i}{r_{2}}\bigg(\frac{\partial}{\partial q_{x}}+i\frac{\partial}{\partial q_{y}}\bigg)e^{i\textbf{q}.\textbf{r}_{2}}.
\end{equation}
Thus we finally get
\begin{equation}
\begin{split}
I_{2}&=\frac{1}{c}\frac{\partial}{\partial q_{z}}\bigg[I(\lambda)+\frac{1}{4}\frac{\partial I(\lambda)}{\partial\lambda}\bigg]\bigg|_{\lambda=1/2},\\
I_{3}&=\frac{1}{c}\bigg[\frac{\partial}{\partial q_{x}}+i\frac{\partial}{\partial q_{y}}\bigg]\bigg[I(\lambda)+\frac{1}{4}\frac{\partial I(\lambda)}{\partial\lambda}\bigg]\bigg|_{\lambda=1/2}.
\end{split}
\end{equation}
The second term in the S-matrix element given in Eq.~(\ref{smatrixp}) is
\begin{equation}
S_{fi}^{(2)}=S_{fi}^{(2),1}+S_{fi}^{(2),2},
\end{equation}
with
\begin{equation}\label{sfi21}
\begin{split}
S_{fi}^{(2),1}&=-\int d\textbf{r}_{2} \frac{\bar{u}(p_{f},s_{f})}{\sqrt{2E_{f}V}}\gamma^{0}\frac{u(p_{i},s_{i})}{\sqrt{2E_{i}V}}\frac{1}{2cp_{B}}\frac{\bar{u}(p_{B},s_{B})\gamma^{0}}{\sqrt{2E_{B}V}}\bigg[\gamma^{0}\frac{E_{B}}{c}-\slashed{p}_{B}\bigg] \varphi^{(+)}_{2S}(\textbf{r}_{2})\\&\times{}_1F_{1}(i\eta_{B}+1,2,i(p_{B}r_{2}+\textbf{p}_{B}.\textbf{r}_{2}))e^{-i\textbf{p}_{B}.\textbf{r}_{2}}(e^{i\mathbf{\Delta}.\textbf{r}_{2}}-1)\\&\times \frac{8\pi^{2}}{\Delta^{2}}\delta(E_{f}+E_{B}-E_{i}-\mathcal{E}_{b}(2S)) e^{\pi \eta_{B}/2} \Gamma(1-i\eta_{B}),
\end{split}
\end{equation}
and
\begin{equation}\label{sfi22}
\begin{split}
S_{fi}^{(2),2}&=-\int d\textbf{r}_{2} \frac{\bar{u}(p_{f},s_{f})}{\sqrt{2E_{f}V}}\gamma^{0}\frac{u(p_{i},s_{i})}{\sqrt{2E_{i}V}}\frac{1}{2c}\frac{\bar{u}(p_{B},s_{B})\gamma^{0}}{\sqrt{2E_{B}V}}\varphi^{'(+)}_{2S}(\textbf{r}_{2}){}_1F_{1}(i\eta_{B}+1,2,i(p_{B}r_{2}+\textbf{p}_{B}.\textbf{r}_{2}))\\&\times e^{-i\textbf{p}_{B}.\textbf{r}_{2}}(e^{i\mathbf{\Delta}.\textbf{r}_{2}}-1)\frac{8\pi^{2}}{\Delta^{2}}\delta(E_{f}+E_{B}-E_{i}-\mathcal{E}_{b}(2S)) e^{\pi \eta_{B}/2} \Gamma(1-i\eta_{B}).
\end{split}
\end{equation}
The operator $[\alpha.\textbf{p}_{B}]$ in Eq.~(\ref{sfi21}) is replaced by $\big[\gamma^{0}\frac{E_{B}}{c}-\slashed{p}_{B}\big]$, and in Eq.~(\ref{sfi22}) $\varphi^{'(+)}_{2S}(\textbf{r}_{2})$ is given by
\begin{equation}
\varphi^{'(+)}_{2S}(\textbf{r}_{2})=[\alpha.\hat{\textbf{r}}_{2}]\varphi^{(+)}_{2S}(\textbf{r}_{2})=\frac{N_{D_{2}}}{4\sqrt{2\pi}}e^{-r_{2}/2}\begin{pmatrix} \frac{i(4-r_{2})}{4c} \\ 0 \\ (2-r_{2})\cos(\theta) \\ (2-r_{2})\sin(\theta)e^{i\phi}\end{pmatrix}.
\end{equation}
$S_{fi}^{(2),1}$ can be recasted in the following form
\begin{equation}\label{sfi21p}
\begin{split}
S_{fi}^{(2),1}&=-[H_{2}(\textbf{q}=\mathbf{\Delta}-\textbf{p}_{B})-H_{2}(\textbf{q}=-\textbf{p}_{B})] \frac{\bar{u}(p_{f},s_{f})}{\sqrt{2E_{f}V}}\gamma^{0}\frac{u(p_{i},s_{i})}{\sqrt{2E_{i}V}}\frac{1}{2cp_{B}}\frac{\bar{u}(p_{B},s_{B})\gamma^{0}}{\sqrt{2E_{B}V}}\\&\times \bigg[\gamma^{0}\frac{E_{B}}{c}-\slashed{p}_{B}\bigg]\frac{8\pi^{2}}{\Delta^{2}}\delta(E_{f}+E_{B}-E_{i}-\mathcal{E}_{b}(2S)) e^{\pi \eta_{B}/2} \Gamma(1-i\eta_{B}),
\end{split}
\end{equation}
where $H_{2}(\textbf{q})$ is the integral expressed by
\begin{equation}\label{h2}
H_{2}(\textbf{q})=\int d\textbf{r}_{2} e^{i\textbf{q}.\textbf{r}_{2}}{}_1F_{1}(i\eta_{B}+1,2,i(p_{B}r_{2}+\textbf{p}_{B}.\textbf{r}_{2}))\varphi^{(+)}_{2S}(\textbf{r}_{2}).
\end{equation}
Replacing the Darwin function in Eq.~(\ref{h2}) by its expression (\ref{Darwin}) leads to
\begin{equation}\label{h2p}
H_{2}(\textbf{q})=\frac{N_{D_{2}}}{4\sqrt{2\pi}}(J_{1},0,J_{2},J_{3})^{T},
\end{equation}
where
\begin{equation}
\begin{split}
J_{1}&=\int d\textbf{r}_{2} e^{i\textbf{q}.\textbf{r}_{2}}e^{-r_{2}/2}(2-r_{2}){}_1F_{1}(i\eta_{B}+1,2,i(p_{B}r_{2}+\textbf{p}_{B}.\textbf{r}_{2})),\\
J_{2}&=\frac{i}{4c}\int d\textbf{r}_{2} e^{i\textbf{q}.\textbf{r}_{2}}e^{-r_{2}/2}(4-r_{2})\cos(\theta){}_1F_{1}(i\eta_{B}+1,2,i(p_{B}r_{2}+\textbf{p}_{B}.\textbf{r}_{2})),\\
J_{3}&=\frac{i}{4c}\int d\textbf{r}_{2} e^{i\textbf{q}.\textbf{r}_{2}}e^{-r_{2}/2}(4-r_{2})\sin(\theta)e^{i\phi}{}_1F_{1}(i\eta_{B}+1,2,i(p_{B}r_{2}+\textbf{p}_{B}.\textbf{r}_{2})).
\end{split}
\end{equation}
To evaluate this three integrals, we introduce a new integral that has been calculated analytically by Attaourti \textit{et al} in \cite{taj}
\begin{equation}\label{IntegralJ}
\begin{split}
J(\lambda)&=\int d\textbf{r} e^{i\textbf{q}.\textbf{r}}\frac{e^{-\lambda r}}{r}{}_1F_{1}(i\eta_{B}+1,2,i(p_{B}r_{2}+\textbf{p}_{B}.\textbf{r}_{2})),\\
&=\frac{4\pi}{(q^{2}+\lambda^{2})}{}_2F_{1}\bigg(i\eta_{B}+1,1,2,-2\frac{(\textbf{q}.\textbf{p}_{B}-i\lambda p_{B})}{q^{2}+\lambda^{2}}\bigg),
\end{split}
\end{equation}
where $\lambda$ is a real variable.\\
In the same way as before and after some manipulations, one gets
\begin{equation}
\begin{split}
J_{1}&=-2\bigg(\frac{\partial J(\lambda)}{\partial\lambda}\bigg)-\frac{\partial^{2} J(\lambda)}{\partial\lambda^{2}}\bigg|_{\lambda=1/2},\\
J_{2}&=\frac{1}{c}\frac{\partial}{\partial q_{z}}\bigg[J(\lambda)+\frac{1}{4}\frac{\partial J(\lambda)}{\partial\lambda}\bigg]\bigg|_{\lambda=1/2},\\
J_{3}&=\frac{1}{c}\bigg[\frac{\partial}{\partial q_{x}}+i\frac{\partial}{\partial q_{y}}\bigg]\bigg[J(\lambda)+\frac{1}{4}\frac{\partial J(\lambda)}{\partial\lambda}\bigg]\bigg|_{\lambda=1/2}.
\end{split}
\end{equation}
For the term $S_{fi}^{(2),2}$, it can be written as
\begin{equation}\label{sfi22p}
\begin{split}
S_{fi}^{(2),2}&=-[H_{3}(\textbf{q}=\mathbf{\Delta}-\textbf{p}_{B})-H_{3}(\textbf{q}=-\textbf{p}_{B})] \frac{\bar{u}(p_{f},s_{f})}{\sqrt{2E_{f}V}}\gamma^{0}\frac{u(p_{i},s_{i})}{\sqrt{2E_{i}V}}\frac{1}{2c}\frac{\bar{u}(p_{B},s_{B})\gamma^{0}}{\sqrt{2E_{B}V}}\\&\times\frac{8\pi^{2}}{\Delta^{2}}\delta(E_{f}+E_{B}-E_{i}-\mathcal{E}_{b}(2S)) e^{\pi \eta_{B}/2} \Gamma(1-i\eta_{B}).
\end{split}
\end{equation}
The quantity $H_{3}(\textbf{q})$ is given by
\begin{equation}\label{h3}
H_{3}(\textbf{q})=\frac{N_{D_{2}}}{4\sqrt{2\pi}}(K_{1},0,K_{2},K_{3})^{T},
\end{equation}
where $K_{1}$, $K_{2}$ and $K_{3}$ are three integrals whose solutions are
\begin{equation}
\begin{split}
K_{1}&=-\frac{i}{c}\bigg[\frac{\partial J(\lambda)}{\partial\lambda}+\frac{1}{4}\frac{\partial^{2} J(\lambda)}{\partial\lambda^{2}}\bigg]\bigg|_{\lambda=1/2},\\
K_{2}&=-i\frac{\partial}{\partial q_{z}}\bigg[2J(\lambda)+\frac{\partial J(\lambda)}{\partial\lambda}\bigg]\bigg|_{\lambda=1/2},\\
K_{3}&=-i\bigg[\frac{\partial}{\partial q_{x}}+i\frac{\partial}{\partial q_{y}}\bigg]\bigg[2J(\lambda)+\frac{\partial J(\lambda)}{\partial\lambda}\bigg]\bigg|_{\lambda=1/2}.
\end{split}
\end{equation}
\subsubsection{Spin-unpolarized TDCS in the SRCBA}
Using the standard procedures of QED \cite{greiner}, we obtain for the spin-unpolarized TDCS
\begin{equation}\label{SRC}
\frac{d\bar{\sigma}^{(SRCBA)}}{dE_{B}d\Omega_{B}d\Omega_{f}}=\frac{1}{16\pi^{3}c^{6}}\frac{|\textbf{p}_{f}||\textbf{p}_{B}|}{|\textbf{p}_{i}|}\frac{e^{\pi\eta_{B}}}{\Delta^{4}}\big|\Gamma(1-i\eta_{B})\big|^{2}\big|\widehat{S}_{fi}^{(1)}+\widehat{S}_{fi}^{(2),1}+\widehat{S}_{fi}^{(2),2}\big|^{2}\bigg|_{E_{f}=E_{i}+\mathcal{E}_{b}(2S)-E_{B}},
\end{equation}
with
\begin{equation}\label{tild1}
\begin{split}
\widehat{S}_{fi}^{(1)}&=\frac{1}{2}\sum_{s_{i},s_{f}}
\sum_{s_{B}}\big[\bar{u}(p_{f},s_{f})\gamma^{0}u(p_{i},s_{i})\big]\big[\bar{u}(p_{B},s_{B})\gamma^{0}\big]\big[i(H_{1}(\textbf{q}=\mathbf{\Delta}-\textbf{p}_{B})-H_{1}(\textbf{q}=-\textbf{p}_{B}))\big],
\end{split}
\end{equation}
\begin{equation}\label{tild2}
\begin{split}
\widehat{S}_{fi}^{(2),1}&=\frac{1}{2}\sum_{s_{i},s_{f}}
\sum_{s_{B}}\big[\bar{u}(p_{f},s_{f})\gamma^{0}u(p_{i},s_{i})\big]\big[\bar{u}(p_{B},s_{B})\gamma^{0}\big]\bigg[\gamma^{0}\frac{E_{B}}{c}-\slashed{p}_{B}\bigg]\frac{1}{2cp_{B}}\\&\times\big[H_{2}(\textbf{q}=\mathbf{\Delta}-\textbf{p}_{B})-H_{2}(\textbf{q}=-\textbf{p}_{B})\big],
\end{split}
\end{equation}
\begin{equation}\label{tild3}
\begin{split}
\widehat{S}_{fi}^{(2),2}&=\frac{1}{2}\sum_{s_{i},s_{f}}
\sum_{s_{B}}\big[\bar{u}(p_{f},s_{f})\gamma^{0}u(p_{i},s_{i})\big]\big[\bar{u}(p_{B},s_{B})\gamma^{0}\big]\frac{1}{2c}\big[H_{3}(\textbf{q}=\mathbf{\Delta}-\textbf{p}_{B})-H_{3}(\textbf{q}=-\textbf{p}_{B})\big].
\end{split}
\end{equation}
In Eq.~(\ref{SRC}), $|\textbf{p}_{i}|$ and $|\textbf{p}_{f}|$ are, respectively, the norms of the initial and final electron momenta.
All the calculations in Eq.~(\ref{SRC}) can be done analytically and only five terms out of nine are nonzero, the diagonal terms $\big|\widehat{S}_{fi}^{(1)}\big|^{2}$, $\big|\widehat{S}_{fi}^{(2),1}\big|^{2}$, $\big|\widehat{S}_{fi}^{(2),2}\big|^{2}$, and $\widehat{S}_{fi}^{(1)\dag}\widehat{S}_{fi}^{(2),1}$, as well as $\widehat{S}_{fi}^{(2),1\dag}\widehat{S}_{fi}^{(1)}$. In Eqs.~(\ref{tild1})-(\ref{tild3}), the different sums over spin states give the following results:
\begin{equation}\label{sum1}
\frac{1}{2}\sum_{s_{i},s_{f}}
\big|\bar{u}(p_{f},s_{f})\gamma^{0}u(p_{i},s_{i})\big|^{2}=2c^{2}\bigg(\frac{2E_{i}E_{f}}{c^{2}}-(p_{i}.p_{f})+c^{2}\bigg),
\end{equation}
\begin{equation}\label{sum2}
\sum_{s_{B}}\bigg|\bar{u}(p_{B},s_{B})\gamma^{0}\bigg[\gamma^{0}\frac{E_{B}}{c}-\slashed{p}_{B}\bigg]\bigg|^{2}=4E_{B}\bigg(\frac{E_{B}^{2}}{c^{2}}-c^{2}\bigg),
\end{equation}
\begin{equation}\label{sum3}
\sum_{s_{B}}\big|\bar{u}(p_{B},s_{B})\gamma^{0}\big|^{2}=4E_{B},
\end{equation}
\begin{equation}\label{sum4}
\frac{1}{2}\sum_{s_{t}}(...)=1(...),
\end{equation}
where $(p_{i}.p_{f})$ in Eq.~(\ref{sum1}) is the scalar product of initial and final four-momentum, and $\sum_{s_{t}}(...)/2$ denotes the averaged sum over the spin states of the target atomic hydrogen.\\
We have to compare the TDCS in Eq.~(\ref{SRC}) with the corresponding one in the Non-Relativistic Coulomb Born Approximation (NRCBA), where the incident and scattered electrons are described by non-relativistic plane waves:
\begin{equation}\label{NRfun}
\psi_{p_{i,f}}(\textbf{r}_{1})=(2\pi)^{-3/2}e^{i\textbf{p}_{i,f}.\textbf{r}_{1}},
\end{equation}
whereas the ejected electron is described by a Coulomb wave function:
\begin{equation}\label{NRCfun}
\psi_{c,p_{B}}(\textbf{r}_{2})=(2\pi)^{-3/2}e^{i\textbf{p}_{B}.\textbf{r}_{2}}e^{\pi/2p_{B}}\Gamma\Big(1+\frac{i}{p_{B}}\Big){}_1F_{1}\Big(-\frac{i}{p_{B}},1,-(p_{B}r_{2}+\textbf{p}_{B}.\textbf{r}_{2})\Big),
\end{equation}
and the hydrogen atomic in its metastable 2S-state is described by the non-relativistic (NR) wave function \cite{joachain}
\begin{equation}
\psi_{2S}^{NR}(\textbf{r}_{2})=\frac{1}{4\sqrt{2\pi}}(2-r_{2})e^{-r_ {2}/2}.
\end{equation}
Thus, the TDCS in the NRCBA is given by:
\begin{equation}\label{NRC}
\frac{d\bar{\sigma}^{(NRCBA)}}{dE_{B}d\Omega_{B}d\Omega_{f}}=\frac{p_{f}p_{B}}{p_{i}}\big|f_{ion}^{CBA}\big|^{2},
\end{equation}
where $f_{ion}^{CBA}$ is the first Coulomb-Born amplitude corresponding to the ionization of metastable 2S-state hydrogen atom by electron impact which is given by:
\begin{equation}
\begin{split}
f_{ion}^{CBA}&=-\frac{2}{\Delta^{2}}\frac{e^{\pi/2p_{B}}}{4(2\pi)^{2}}\Gamma\bigg(1-\frac{i}{p_{B}}\bigg)\bigg[-2\frac{\partial I(\textbf{q}=\mathbf{\Delta}-\textbf{p}_{B})}{\partial\lambda}-\frac{\partial^{2} I(\textbf{q}=\mathbf{\Delta}-\textbf{p}_{B}}{\partial \lambda^{2}}\bigg]\bigg|_{\lambda=1/2},
\end{split}
\end{equation}
where the integral $I(\textbf{q})$ is the same as that given previously in Eq.~(\ref{IntegralI}), but here $\eta_{B}=1/p_{B}$.
\subsection{Symmetric coplanar geometry}
The symmetric coplanar geometry can be considered as a particular case of asymmetric coplanar geometry. Let us first remind that the symmetric geometry, also called binary geometry, is defined by the requirement that the kinetic energies of the scattered and ejected electrons are nearly the same, and the scattered and ejected electron angles with respect to the incident beam direction are equal to each other.
In this section, we present the relativistic formalism of the $(e,2e)$ reaction in the Relativistic Plane Wave Born Approximation (RPWBA), where the incident, scattered, and ejected electrons are described by relativistic plane
waves, and the hydrogen atom in its metastable 2S-state is described by the relativistic exact function given by:
\begin{equation}
\phi_{i}(t,\textbf{r}_{2})=\exp[-i\mathcal{E}_{b}(2S) t]\varphi^{(\pm),2S}_{Exact}(\textbf{r}_{2}),
\end{equation}
where $\mathcal{E}_{b}(2S)$ is the binding energy of the metastable 2S-state of atomic hydrogen given in (\ref{binding}). For spin up, $\varphi^{(+),2S}_{Exact}(\textbf{r}_{2})$ is expressed by:
\begin{equation}\label{Exact}
\begin{split}
\varphi^{(+),2S}_{Exact}(\textbf{r}_{2})&=\frac{1}{2\sqrt{4\pi}}\frac{(2Z)^{\gamma_{H}+1/2}}{a_{2S}^{\gamma_{H}+1}}\sqrt{\frac{2\gamma_{H}+1}{(a_{2S}+1)\Gamma(2\gamma_{H}+1)}}r_{2}^{\gamma_{H}-1}e^{-Zr_{2}/a_{2S}}\\&\times\begin{pmatrix} ig_{2S_{1/2}}(r_{2}) \\ 0 \\ f_{2S_{1/2}}(r_{2})\cos(\theta) \\ f_{2S_{1/2}}(r_{2})\sin(\theta)e^{i\phi}\end{pmatrix},
\end{split}
\end{equation}
where $\theta$ and $\phi$ are the spherical coordinates of $\textbf{r}_{2}$. The two quantities $g_{2S_{1/2}}(r_{2})$ and $f_{2S_{1/2}}(r_{2})$ are such as:
\begin{equation}
\begin{split}
g_{2S_{1/2}}(r_{2})&=\sqrt{1+\frac{Z\alpha}{\sqrt{2(1-\gamma_{H})}}}\bigg[\bigg(1-\frac{2Zr_{2}}{a_{2S}(2\gamma_{H}+1)}\bigg)\big(a_{2S}+1\big)-1\bigg],\\
f_{2S_{1/2}}(r_{2})&=\sqrt{1-\frac{Z\alpha}{\sqrt{2(1-\gamma_{H})}}}\bigg[\bigg(1-\frac{2Zr_{2}}{a_{2S}(2\gamma_{H}+1)}\bigg)\big(a_{2S}+1\big)+1\bigg],
\end{split}
\end{equation}
where $Z$ is the atomic number, and the two parameters $\gamma_{H}$ and $a_{2S}$ are given by:
\begin{equation}
\begin{split}
\gamma_{H}&=\sqrt{1-Z^{2}\alpha^{2}},\\
a_{2S}&=\sqrt{2(\gamma_{H}+1)},
\end{split}
\end{equation}
with $\alpha=1/c$ is the fine structure constant.\\
Substituting all these expressions into the first Born S-matrix element (\ref{smatrix}) and after some manipulations, one gets
\begin{equation}\label{RPWBA}
\begin{split}
\frac{d\bar{\sigma}^{(RPWBA)}}{dE_{B}d\Omega_{B}d\Omega_{f}}&=\frac{1}{2}\frac{p_{f}p_{B}}{c^{6}p_{i}\Delta^{4}}\bigg(\frac{1}{2}\sum_{s_{i},s_{f}}
\big|\bar{u}(p_{f},s_{f})\gamma^{0}u(p_{i},s_{i})\big|^{2}\bigg)\sum_{s_{B}}\big|\bar{u}(p_{B},s_{B})\gamma^{0}\big|^{2}\\&\times\big|\Phi_{2,1/2,1/2}(\textbf{q}=\mathbf{\Delta}-\textbf{p}_{B})-\Phi_{2,1/2,1/2}(\textbf{q}=-\textbf{p}_{B})\big|^{2}.
\end{split}
\end{equation}
The different sums over spin states $s_{i}$, $s_{f}$ and $s_{B}$  are given before in Eqs.~(\ref{sum1}-\ref{sum3}). The functions $\Phi_{2,1/2,1/2}(\textbf{q})$ are the Fourier transforms of the relativistic atomic hydrogen wave functions
\begin{equation}
\Phi_{n=2,j=1/2,m=1/2}(\textbf{q})=(2\pi)^{-3/2}\int d\textbf{r}_{2}e^{i\textbf{q}.\textbf{r}_{2}}\varphi^{(+),2S}_{Exact}(\textbf{r}_{2}),
\end{equation}
and $\mathbf{\Delta}=\textbf{p}_{i}-\textbf{p}_{f}$ is the momentum transfer. Replacing the exact function $\varphi^{(+),2S}_{Exact}(\textbf{r}_{2})$ by its expression (\ref{Exact}) yields
\begin{equation}
\begin{split}
\Phi_{n=2,j=1/2,m=1/2}(\textbf{q})&=(2\pi)^{-3/2}\frac{1}{2\sqrt{4\pi}}\frac{(2Z)^{\gamma_{H}+1/2}}{a_{2S}^{\gamma_{H}+1}}\sqrt{\frac{2\gamma_{H}+1}{(a_{2S}+1)\Gamma(2\gamma_{H}+1)}}\\&\times
\begin{pmatrix} \int d\textbf{r}_{2}e^{i\textbf{q}.\textbf{r}_{2}} r_{2}^{\gamma_{H}-1}e^{-Zr_{2}/a_{2S}}ig_{2S_{1/2}}(r_{2}) \\ 0 \\ \int d\textbf{r}_{2}e^{i\textbf{q}.\textbf{r}_{2}} r_{2}^{\gamma_{H}-1}e^{-Zr_{2}/a_{2S}}f_{2S_{1/2}}(r_{2})\cos(\theta) \\ \int d\textbf{r}_{2}e^{i\textbf{q}.\textbf{r}_{2}} r_{2}^{\gamma_{H}-1}e^{-Zr_{2}/a_{2S}}f_{2S_{1/2}}(r_{2})\sin(\theta)e^{i\phi}\end{pmatrix}.
\end{split}
\end{equation}
The expression of the TDCS in the Semi-Relativistic Plan Wave Born Approximation (SRPWBA) remains similar to that given in the RPWBA (\ref{RPWBA}), except the expression of the Fourier transform which changes since the wave function describing the hydrogen atom in the SRPWBA is replaced by the Darwin wave function that we have previously expressed in Eq.~(\ref{Darwin}). This TDCS in Eq.~(\ref{RPWBA}) is to be compared with the corresponding one in the Non-Relativistic Plane Wave Born Approximation (NRPWBA), where the incident, scattered, and ejected electrons are described by non-relativistic plane waves:
\begin{equation}\label{NRPWBA}
\begin{split}
\frac{d\bar{\sigma}^{(NRPWBA)}}{dE_{B}d\Omega_{B}d\Omega_{f}}&=\frac{2^{10}}{\pi^{2}\Delta^{4}}\frac{p_{f}p_{B}}{p_{i}}\bigg[\frac{4\textbf{q}^{2}-1}{(1+4\textbf{q}^{2})^{3}}-\frac{4\textbf{q}_{0}^{2}-1}{(1+4\textbf{q}^{2}_{0})^{3}}\bigg]^{2},
\end{split}
\end{equation}
where $\textbf{q}=\mathbf{\Delta}-\textbf{p}_{B}$ and $\textbf{q}_{0}=-\textbf{p}_{B}$.
\section{Results and discussion}
In this paper, we develop an exact relativistic model, in the first Born approximation, to study the ionization of the metastable 2S-state hydrogen atom by electron impact at high energies in the asymmetric and symmetric coplanar geometries. The required derivatives of hypergeometric functions and all integrals resulting from the Fourier transforms of the relativistic and semirelativistic atomic hydrogen wave functions are computed in closed analytic forms using the programming language MATHEMATICA, which is also used to plot the various figures of the present work.
In this section, we will present all the numerical results obtained in both asymmetric and symmetric geometries; during that, we will follow the same arrangement that we adopted in the previous section. We will start first with the results obtained in the case of asymmetric geometry and then symmetric geometry. All the TDCSs are given in atomic units.
\subsection{Asymmetric coplanar  geometry}
\begin{figure}[h!]
	\centering
	\includegraphics[width=11cm,height=9cm]{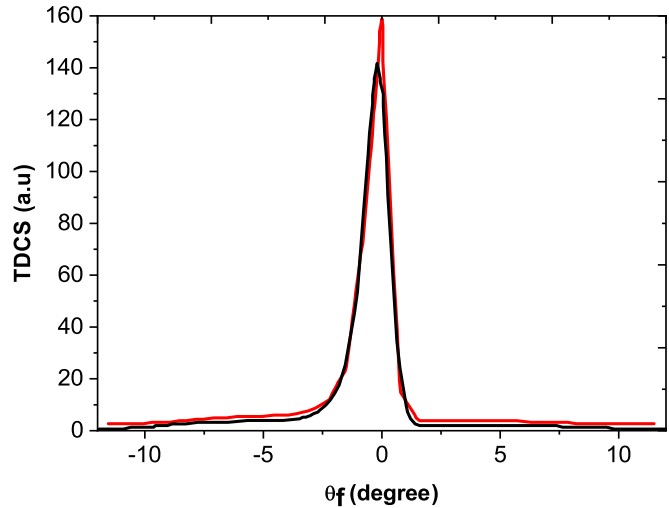}
	\caption{The TDCS of the (e, 2e) ionization of hydrogen 2S in terms of the scattering angle $\theta_{f}$. The incident and the ejected electron kinetic energies are $250$ eV and $5$ eV respectively and the ejection angle $\theta_{B}=20^{\circ}$. The solid red line gives our results (NRCBA) given in (\ref{NRC}) and the solid black line those of Hafid \textit{et al.} \cite{hafid}}\label{comparison20}
\end{figure}

We will begin our discussion, in this case, by comparing our results with those obtained by Hafid \textit{et al.} \cite{hafid} in the nonrelativistic domain. Hafid's results were obtained using the well-known approximation BBK model of Brauner \textit{et al.} \cite{BBK}, and when Hafid presented his results, he also compared with those obtained by Coulomb wave function and second born calculations of Vucic \textit{et al.} \cite{vucic} with respect to the incoming electron kinetic energy of $250$ eV and the ejected electron kinetic energy of $5$ eV. In the following figures (Fig.~(\ref{comparison3}), Fig.~(\ref{comparison5}) and Fig.~(\ref{comparison20})), which contain the comparison with other theoretical calculations, the angular choice is as follows: $p_{i}$ is along the $z$-axis and $\theta_{i}=0^{\circ}, \phi_{i}=0^{\circ}$. For the scattered electron, we
choose $\phi_{f}=0^{\circ}$ and $\theta_{f}$ is fixed in Figs.~(\ref{comparison3}) and (\ref{comparison5}), respectively, to the values $\theta_{f}=3^{\circ}$ and  $\theta_{f}=5^{\circ}$, while in Fig.~(\ref{comparison20}) $\theta_{f}$ varies from $-12^{\circ}$ to $12^{\circ}$. For the ejected electron, we choose $\phi_{B}=0^{\circ}$ and $\theta_{B}$ varies from $0^{\circ}$ to $360^{\circ}$ in Figs.~(\ref{comparison3}) and (\ref{comparison5}) and it is fixed to the value $\theta_{B}=20^{\circ}$ in Fig.~(\ref{comparison20}).
\begin{figure}[h!]
	\centering
	\includegraphics[width=11cm,height=9cm]{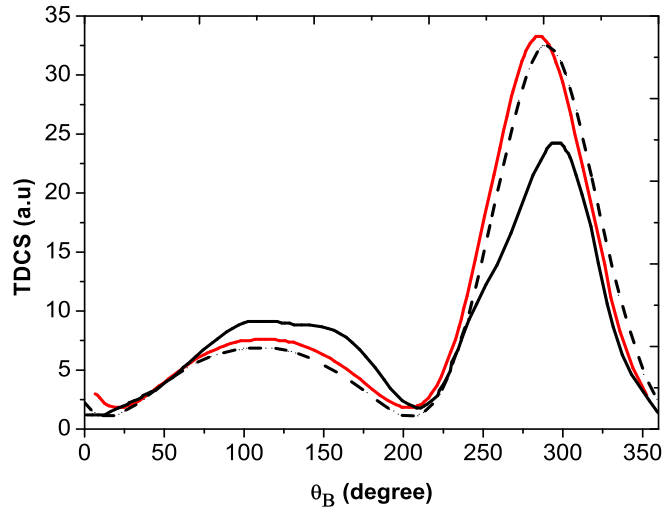}
	\caption{The TDCS of the (e, 2e) ionization of hydrogen 2S in terms of the ejection angle $\theta_{B}$. The incident and the ejected electron kinetic energies are $250$ eV and $5$ eV respectively and the scattering angle $\theta_{f}= 3^{\circ}$. The solid red line gives our results (NRCBA) given in Eq.(\ref{NRC}), the solid black line those of Hafid \textit{et al.} \cite{hafid} and the dashed line results obtained by Coulomb wave function.}\label{comparison3}
\end{figure}\\
We compare, in Fig~(\ref{comparison3}), our results in the NRCBA (Eq.(\ref{NRC})) with those of Hafid \textit{et al.} and those obtained by Coulomb wave function (where a Coulomb wave is used for the ejected electron and plane waves for the incident and scattered electrons: this model is called 1CW,
one coulomb wave) for the incident kinetic energy of $250$ eV, ejection kinetic energy value of $5$ eV and  the scattering angle of $\theta_{f}=3^{\circ}$. Our results in the NRCBA model were obtained, as we have seen in the theoretical calculations in the previous section, by using a Coulomb wave function to describe only the ejected electron, whereas the fast incident and scattered electrons are described by non-relativistic plane wave functions, thus neglecting the Coulomb interaction of the fast scattered electron with the system. Thus, our NRCBA model is the same as the well-known 1CW model. Therefore, the results obtained in both models should be compatible with each other. As we can see in Fig.~{\ref{comparison3}}, the two curves representing these results (red and dashed curves
respectively) have good agreement and convergence. They are very close, both in the shape of the curve and the location of the peaks, as well as in the order of magnitude. These two results obtained using the Coulomb wave functions remain different in magnitude, as well as in the height of the binary peak from the result obtained by Hafid \textit{et al.} using BBK approximation.
\begin{figure}[h!]
	\centering
	\includegraphics[width=11cm,height=9cm]{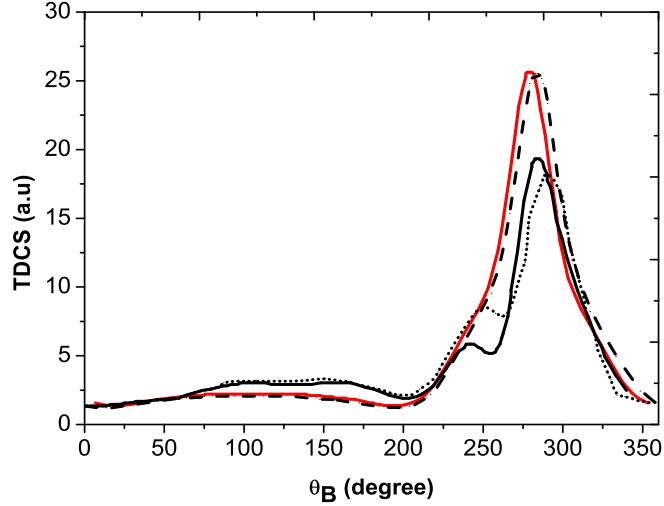}
	\caption{The TDCS of the (e, 2e) ionization of hydrogen 2S in terms of the ejection angle $\theta_{B}$. The incident and the ejected electron kinetic energies are $250$ eV and $5$ eV respectively and the scattering angle $\theta_{f}= 5^{\circ}$. The solid red line gives our results (NRCBA) given in (\ref{NRC}), the solid black line those of Hafid \textit{et al.} \cite{hafid}, the dotted line those of Vucic \textit{et al.} \cite{vucic} and the  dashed line results obtained using Coulomb wave function.}\label{comparison5}
\end{figure}
Figure (\ref{comparison5}) represents similar parameterization  as in Fig.~(\ref{comparison3}), but with the scattering angle $\theta_{f}= 5^{\circ}$. We have also included here the second Born results of Vucic \textit{et al.} \cite{vucic}. Again, it is clearly seen from this figure, that our model NRCBA still gives the same results compared to the  model 1CW with an apparent difference between them and the results obtained from other calculations. In the recoil region, all results remain close in form and magnitude. It is interesting to see that the results of Hafid \textit{et al.} reveal a peak which is present also in the second Born calculation of Vucic \textit{et al.} and absent in the calculation used in the first Born Coulomb approximation (the ejected electron is described by a Coulomb wave and the incident and scattered electrons by plane waves). Comparing Figs.~(\ref{comparison3}) and (\ref{comparison5}), we note that the magnitude of the two peaks decreases with increasing the scattering angle $\theta_{f}$.
We study in Fig.~(\ref{comparison20}), for the same kinetic energy, the variation of the NRCBA in terms of the scattering angle $\theta_{f}$ for the ejection angle $\theta_{B}=20^{\circ}$. The comparison with the results of Hafid \textit{et al.} is also included. This actually gives a sharp peak,
higher than the other peaks in the previous figures. We observe also that the scattered electron, which is relatively faster than the ejected one, goes out with small angles. The discrepancies among the results obtained using the Coulomb wave functions and other theoretical results obtained from BBK and second Born approximations shown in Figs.~(\ref{comparison3}) and (\ref{comparison5}) are expected due to the difference of the approaches used in each theoretical study. The final state wave function in the BBK results of Hafid \textit{et al.} is improved by including the effects of all long range Coulomb interactions and the repulsion between the two final electrons, leading to results comparable to the second Born approximation (Vucic \textit{et al.}). It has been proved, for the ionization of the hydrogen (1S) atom by electron impact, that it is the BBK model which gives the excellent agreement with experiment at impact
energies greater than $150$ eV \cite{BBK}. From a theoretical point of view, the difficulty resides in the description of the three-body final state (residual ion, scattered and ejected electrons) in Coulomb interaction. The question of the correlations of the different wave functions involved in its description remains, nowadays, an unanswered question. Many attempts have been made to determine the TDCS by neglecting the interaction of the fast scattered electron with the system or by using a product of two Coulomb wave functions for the scattered and ejected electrons. Only an experimental benchmark study could judge the validity or not of a theoretical approach, and therefore our only hope is in future experiments that will give us the opportunity to validate our results at different energies.
\begin{figure}[h!]
	\centering
	\includegraphics[scale=0.65]{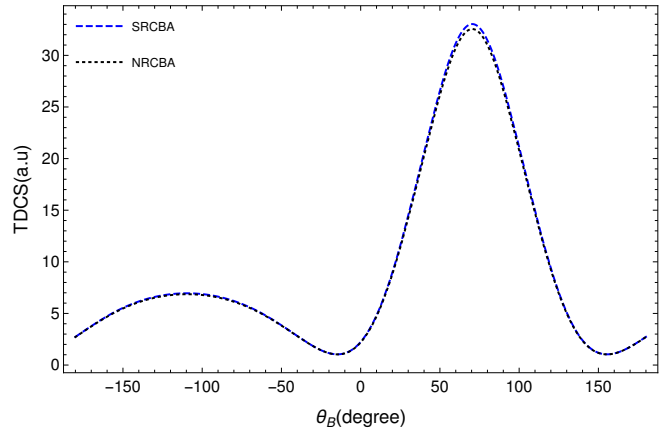}
	\caption{The two TDCSs as a function of the ejection angle $\theta_{B}$. The incident and the ejected electron kinetic energies are $250$ eV and $5$ eV respectively and the scattering angle $\theta_{f}=3^{\circ}$. The other angles are chosen as follows: $\theta_{i}=0^{\circ}, \phi_{i}=0^{\circ}$, $\phi_{f}=0^{\circ}$ and $\phi_{B}=180^{\circ}$.}\label{recoilbinary}
\end{figure}
Figure (\ref{recoilbinary}) depicts the TDCS in the SRCBA and the corresponding one in the NRCBA for the scattering angle $\theta_{f}=3^{\circ}$. The incident electron kinetic energy is $T_{i}=250$ eV and the ejected electron kinetic energy is $T_{B}=5$ eV. We see, as in the case of the ground state \cite{taj}, that the two curves are identical and have two peaks, one in the interval between $-180^{\circ}$ and $0^{\circ}$ (recoil peak) due to projectile-nucleus interaction and the other in the range between $0^{\circ}$ and $180^{\circ}$ (binary peak) due to the electron-electron interaction. The fact that the SRCBA gives the same results compared to NRCBA at low energies can be considered as a criterion for checking its consistency and its validity. However, even in the nonrelativistic regime, small effects are presented; due to the semirelativistic treatment of the wave functions that we have used in the SRCBA, and these can only be related to the spin effect. 
\subsection{Symmetric coplanar  geometry}
\begin{figure}[h!]
	\centering
	\includegraphics[scale=0.6]{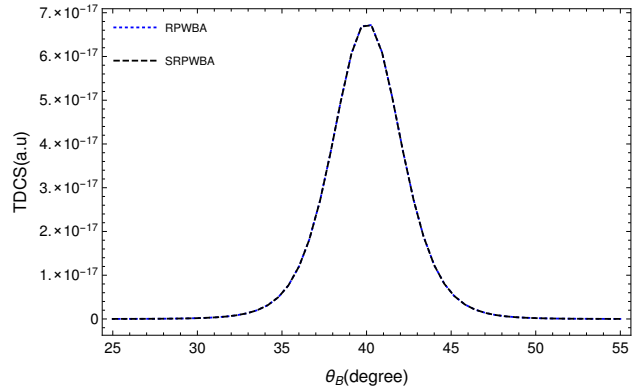}
	\caption{The two TDCSs in the relativistic regime as a function of the ejection angle $\theta_{B}$. The incident and the ejected electron kinetic energies are $511002$ eV and $255499.3$ eV respectively and the scattering angle $\theta_{f}= 45^{\circ}$.}\label{RandSRconfondue}
\end{figure}

In symmetric geometry, as we mentioned previously, the kinetic energies of both scattered and ejected electrons are required to be approximately equal. The TDCS, in all models studied in the previous section, depends explicitly on the kinetic energy values of the scattered and ejected electrons, in addition to the different spherical coordinates related to each electron. Therefore, care must be taken when choosing the values of these kinetic energies, so that the above-mentioned geometry condition is fulfilled. We remind the reader here of the relation that allows us to obtain these values without violating the requirement of symmetric geometry. Using the kinetic energy conservation $T_{f}=T_{i}+\varepsilon_{2S}-T_{B}$, we find that, according to the condition $T_{f}=T_{B}$, $T_{B}=(T_{i}+\varepsilon_{2S})/2$, where $\varepsilon_{2S}=-3.4$ eV $=-0.125$ a.u. is the nonrelativistic binding energy of atomic hydrogen in its metastable 2S-state. Thus, every kinetic energy of the incoming electron corresponds to a kinetic energy of the scattered electron determined from that relation so that the condition of symmetric geometry always remains true. For the symmetric coplanar geometry, we choose the following angular situation where $p_{i}$ is along the $z$-axis ($\theta_{i}=0^{\circ}, \phi_{i}=0^{\circ}$). For the scattered electron, we choose ($\theta_{f}=45^{\circ}, \phi_{f}=0^{\circ}$) and for the ejected electron we choose $\phi_{B}=180^{\circ}$ and the angle $\theta_{B}$ varies differently from a figure to another. First of all, we will try to clarify the limit between the relativistic and non-relativistic domains in the case of the ionization of the hydrogen atom from its metastable 2S-state. Because, compared to the results of the ground state, we found that there is a significant difference between the two non-relativistic limit values. If the hydrogen atom is ionized from its ground 1S-state, the non-relativistic limit value is defined by the relativistic parameter ($\gamma=[1-(\beta/c)^{2}]^{-1/2}$) value of $1.0053$ which corresponds to an incident electron kinetic energy of $2700$ eV \cite{taj}. We recall here that, in atomic units, the kinetic energy is related to $\gamma$ parameter by the following relation: $T_{i}=c^{2}(\gamma-1)$. It means that when the value of the relativistic parameter $\gamma$ is greater than $1.0053$, a difference between the relativistic and non-relativistic kinetic energies will appear. In the case of the ionization of the hydrogen atom from its metastable 2S-state, we found that the non-relativistic limit changed and increased slightly from $2700$ eV until it reached the value of $4250$ eV.
\begin{figure}[h!]
	\centering
	\includegraphics[scale=0.6]{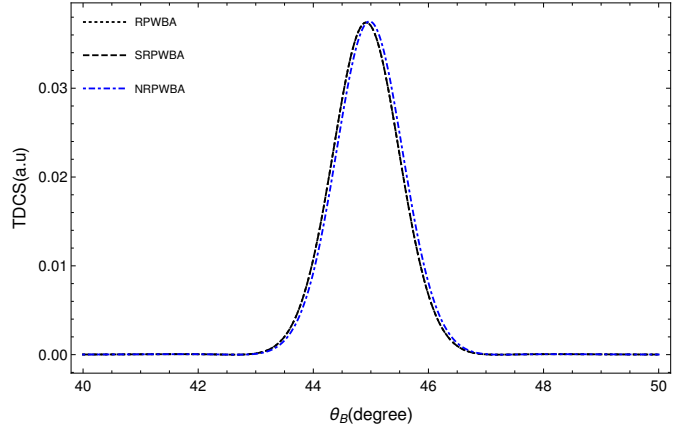}
	\caption{The Three TDCSs of the (e, 2e) ionization of hydrogen 2S as a function of the ejection angle $\theta_{B}$. The incident and the ejected electron kinetic energies are $4250$ eV and $2123.3$ eV respectively and the scattering angle $\theta_{f}= 45^{\circ}$.}\label{limiteNR}
\end{figure}
\begin{figure}[h!]
	\subfloat[]{\label{3TDCS10k}\includegraphics[height=5.5cm,width=.5\linewidth]{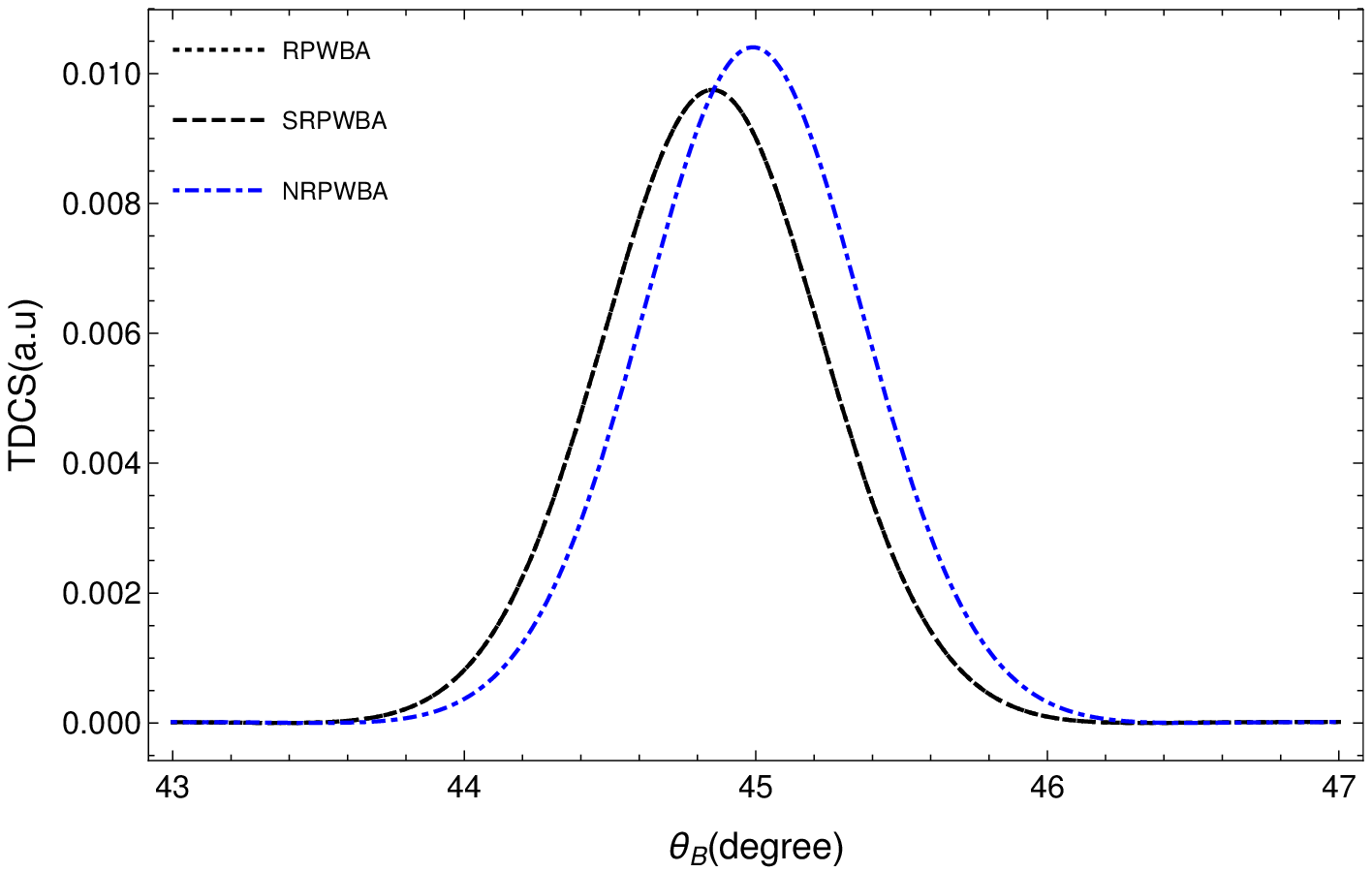}}\hspace*{.5cm}
	\subfloat[]{\label{3TDCS20k}\includegraphics[height=5.5cm,width=.5\linewidth]{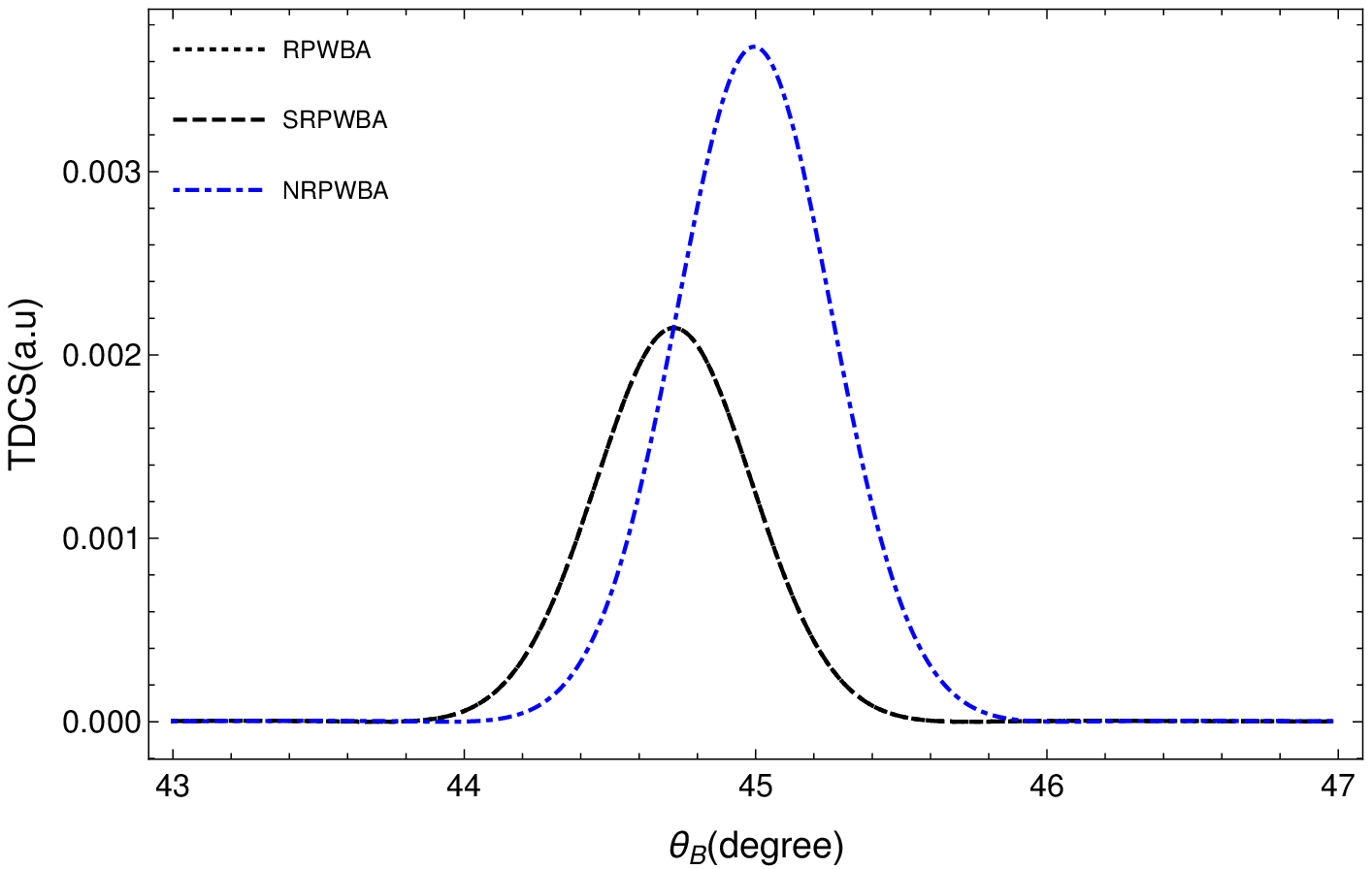}}
	\caption{Same as in Fig.~(\ref{limiteNR}) but for the incident and the ejected electron kinetic energies of (a) $10000$ eV and $4998.3$ eV and (b) $20000$ eV and $9998.3$ eV respectively.}
	\label{10k20k}
\end{figure}
In Fig.~(\ref{limiteNR}), it can be seen that there is no difference at all between the TDCSs (RPWBA, SRPWBA and
NRPWBA) in the non-relativistic limit, since all the curves of the three TDCSs are almost equal and identical. This figure represents the first check of our models in particular in the non-relativistic limit ($T_{i}=4250$ eV, $T_{f}=T_{B}=2123.3$ eV). But, we note that when we pass this limit by raising the kinetic energy of the incoming electron to $10$ keV and $20$ keV, the non-relativistic TDCS begins to differ from the other two TDCSs that remain equal as depicted in Fig.~(\ref{10k20k}). Thus, the agreement between the relativistic and
nonrelativistic models is good from the nonrelativistic
limit and below ($T_{i}\leq 4250$ eV), but the disagreement increases at high energies. It appears from Fig.~(\ref{10k20k}) that at the relativistic domain, the effects of the spin terms and the relativity begin to be noticeable and that the non-relativistic formalism is no longer valid. We notice from Fig.~(\ref{limiteNR}) that there is a parfait symmetry around the value $\theta_{B}=45^{\circ}$ and the three TDCSs are all peaked in the vicinity of the same value. We also note from Fig.~(\ref{10k20k}) that the binary peak position in the relativistic domain begins with a shift towards smaller values than $45^{\circ}$. Comparing Figs.~(\ref{limiteNR}) and (\ref{10k20k}), it is clearly seen that the magnitude of the binary peak decreases with increasing the kinetic energy of the incident electron, which is the usual behavior in charged particle-impact
ionization of an atom. By the way, these two relativistic and semirelativivstic TDCSs (RPWBA and SRPWBA) remain the same and equal, regardless of the kinetic energy value of the incoming electron. For example, we give in Fig.~(\ref{RandSRconfondue}) a representation of the RPWBA and SRPWBA at high incident kinetic energy of $511002$ eV. It appears to us through Fig.~(\ref{RandSRconfondue}) that the two TDCSs (RPWBA and SRPWBA), despite the different wave functions used to describe the hydrogen atom in each of them, give the same results even at high energies. This fact was proven and applied in more than one place when studying the excitation or ionization of the hydrogen atom where it is sufficient to use only the Darwin wave function, instead of the exact analytical wave function, as a semirelativistic state to represent the atomic hydrogen, and it was found that this gives nearly the same results as the
exact description only when the condition $Z\alpha\ll 1$ is fulfilled. This is precisely the reason why, when studying theoretically asymmetric geometry in the previous section, we were satisfied with only the treatment of the SRCBA model without the corresponding one in the Relativistic Coulomb Born Approximation (RCBA), so there is no need to complicate the calculation more as long as both give the same results.
\begin{figure}[h!]
	\centering
	\includegraphics[scale=0.55]{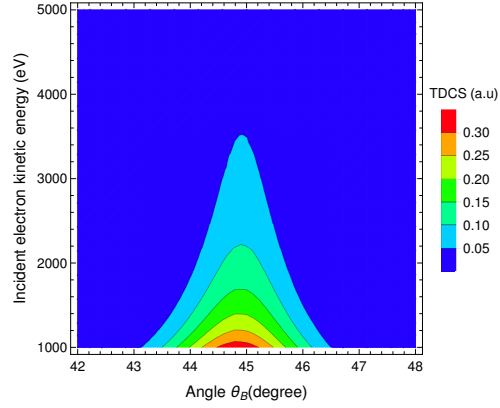}
	\caption{The TDCS, in RPWBA, as a function of the angle $\theta_{B}$ of the ejected electron and the incident electron kinetic energy $T_{i}$ varying from $1000$ eV to $5000$ eV. We have used the condition $40^{\circ}\leq\theta_{f}=\theta_{B}\leq50^{\circ}$.}\label{curve3d1}
\end{figure}
For the sake of illustration, in a similar way to the 2D-plot, the contour plot in Fig.~(\ref{curve3d1}) exhibits more information on the variation and the shape of the TDCS in the RPWBA versus both incident electron kinetic energy and angle $\theta_{B}$ in the binary coplanar geometry.  For the variation with respect to $\theta_{B}$, we observe that the TDCS decreases at small and large angles. We see also that the TDCS presents a maximum only at the particular point of $\theta_{B}=\theta_{f}=45^{\circ}$, and its magnitude at this particular point decreases as the electron kinetic energy increases.
\begin{figure}[h!]
	\centering
	\includegraphics[scale=0.55]{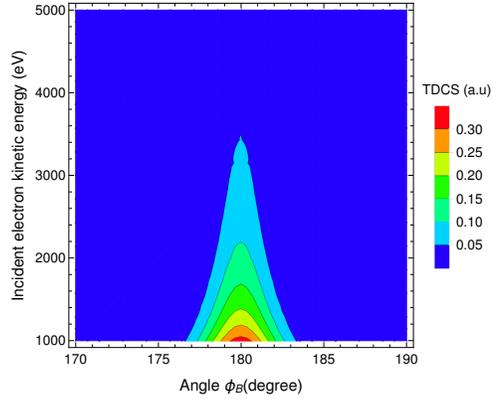}
	\caption{The TDCS, in RPWBA, as a function of the angle $\phi_{B}$ of the ejected electron and the incident electron kinetic energy $T_{i}$ varying from $1000$ eV to $5000$ eV for $\theta_{B}=\theta_{f}=45^{\circ}$.}\label{curve3d2}
\end{figure}
Figure (\ref{curve3d2}) shows that when the incident electron kinetic energy increases provided that $\theta_{B}=\theta_{f}=45^{\circ}$, the peak of the TDCS decreases and remains nearly around $\phi_{B}=180^{\circ}$. From Fig.~(\ref{curve3d2}), we see that as the energy increases, the probability to observe the ejected electron in the direction $\phi_{B}=180^{\circ}$ diminishes progressively. Figure (\ref{curve3d2}) also shows that for all figures in the symmetric geometry, in which we choose $\phi_{B}$ to be constant, it must be equal to $180^{\circ}$
since the pick is clearly located at the same value.
\begin{figure}[h!]
	\centering
	\includegraphics[scale=0.6]{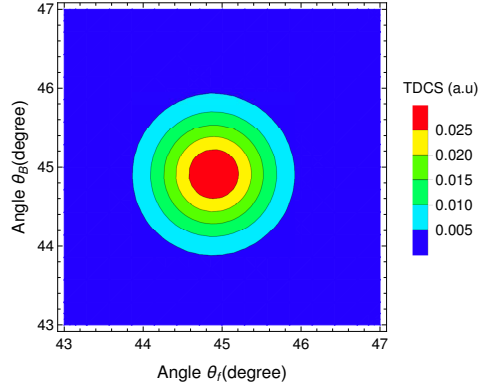}
	\caption{The TDCS, in RPWBA, as a function of the angle $\theta_{B}$ of the ejected electron and the angle $\theta_{f}$ of the scattered electron for $T_{i}=5000$ eV and  $T_{B}=2498.3$ eV.}\label{contourplot3}
\end{figure}
Figure (\ref{contourplot3}) represents the variations of TDCS in the RPWBA in terms of the scattered and ejected electron angles at the energies $T_{i}=5000$ eV and  $T_{B}=2498.3$ eV. The purpose of including this figure is to show how important the condition on both angles to be verified  in the symmetric coplanar geometry. Through this figure, it becomes clear to us that the TDCS represents a maximum value at $\theta_{f}=\theta_{B}=45^{\circ}$ and begins to decrease directly in the areas where this condition is broken. From here it becomes evident that we must always respect this requirement and take it into account when working within the symmetric coplanar geometry.
\begin{figure}[h!]
	\centering
	 \subfloat[]{\label{SRCandNRCdiff10}\includegraphics[height=5cm,width=.4\linewidth]{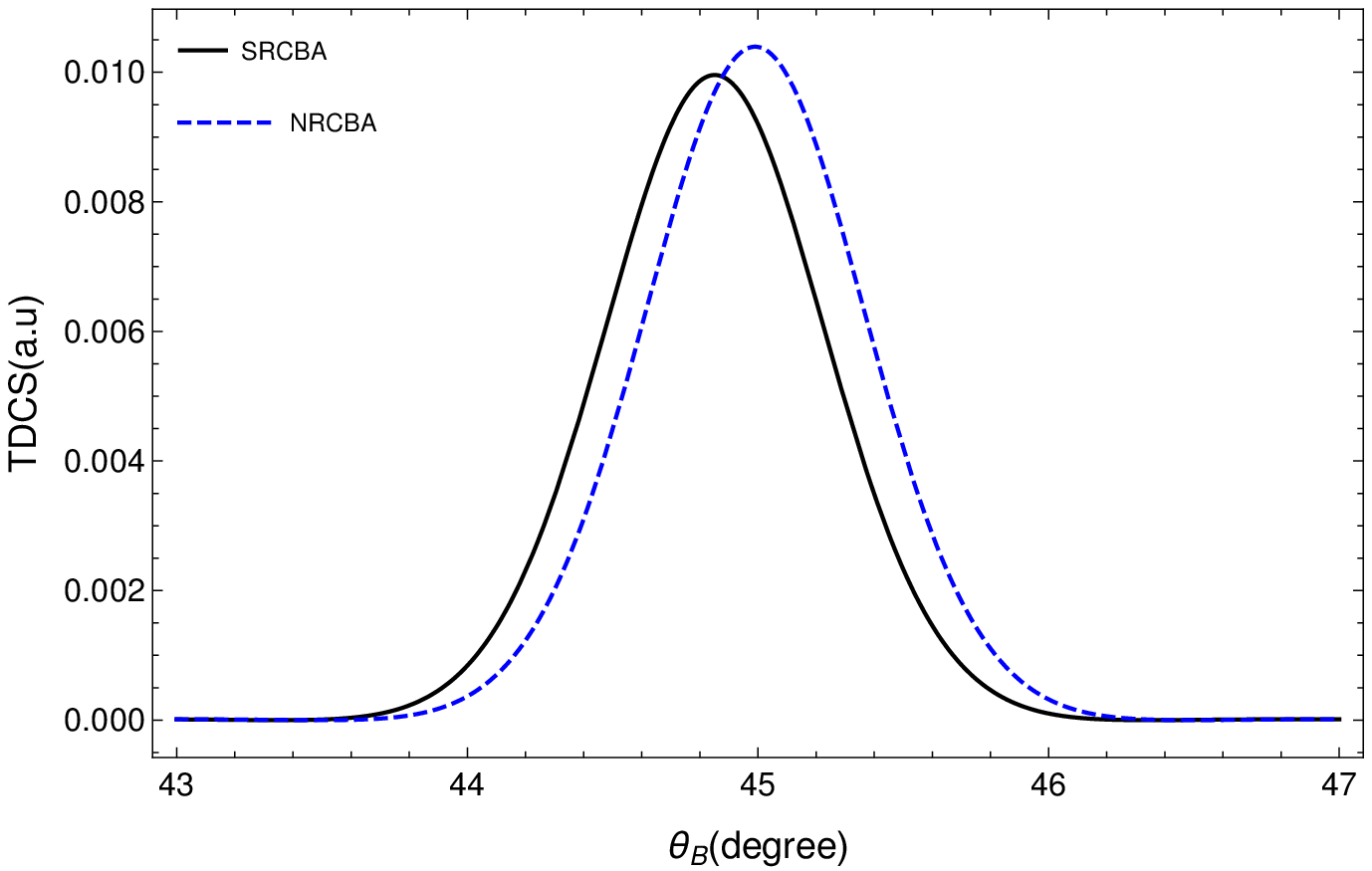}}\hspace*{.5cm}
	 \subfloat[]{\label{SRCandNRCdiff15}\includegraphics[height=5cm,width=.4\linewidth]{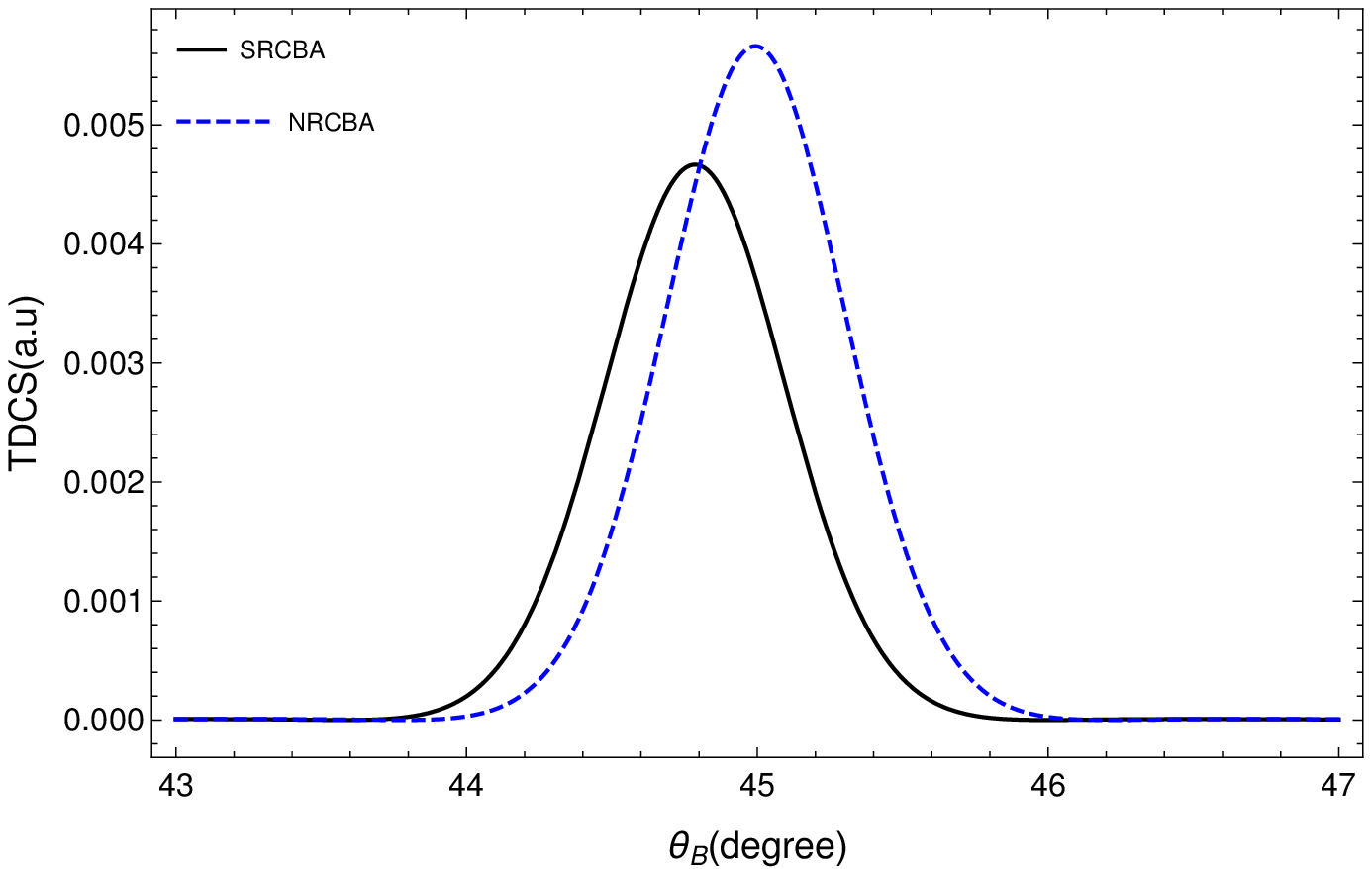}}\\
	\subfloat[] {\label{SRCandNRCdiff20}\includegraphics[height=5cm,width=.4\linewidth]{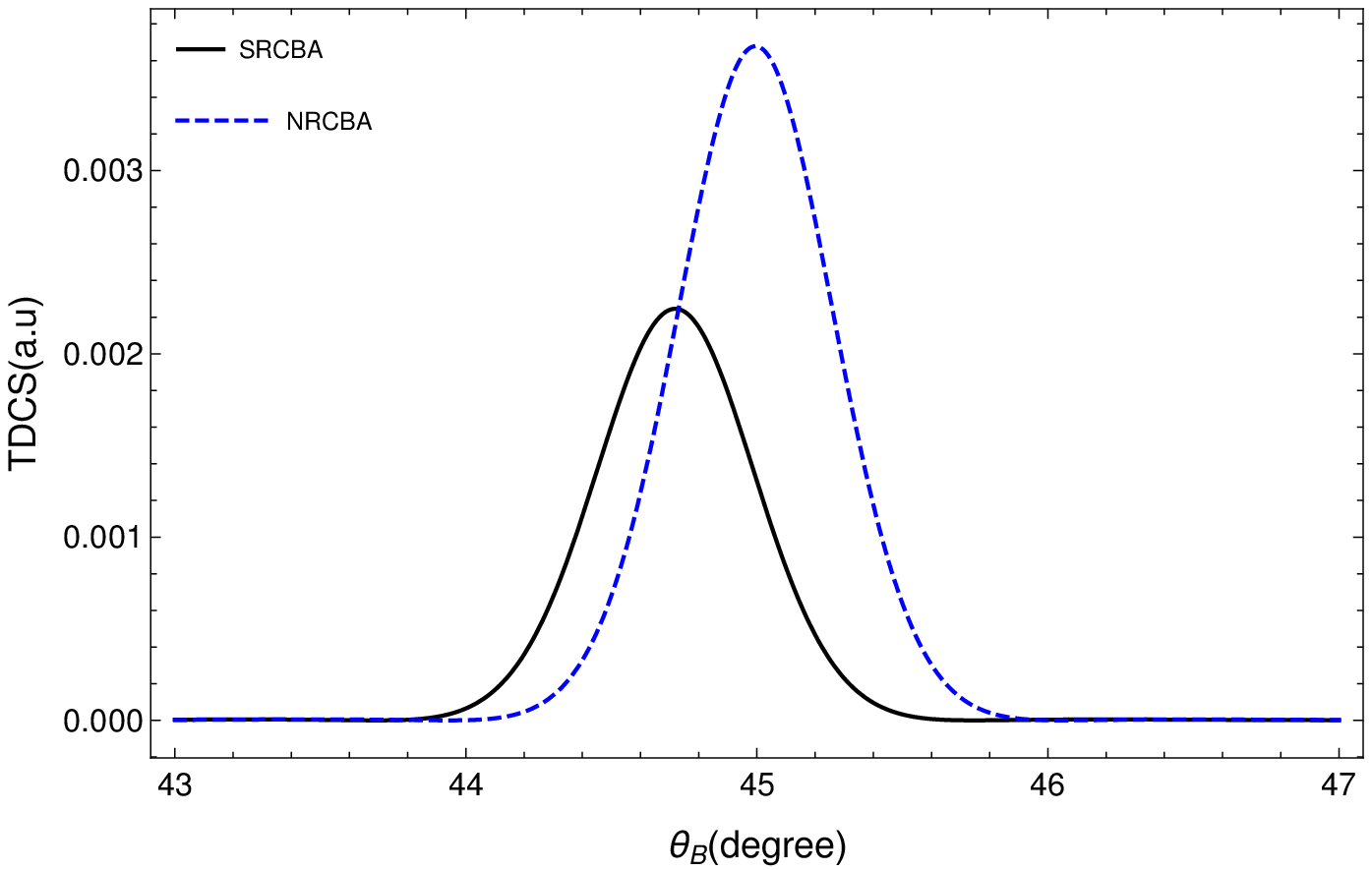}}
	\caption{The two TDCSs in symmetric coplanar geometry as a function of the ejection angle $\theta_{B}$ for the scattering angle $\theta_{f}=45^{\circ}$. The incident and the ejected electron kinetic energies are (a) $T_{i}=10000$ eV and $T_{B}=4998.3$ eV, (b) $T_{i}=15000$ eV and $T_{B}=7498.3$ eV and (c) $T_{i}=20000$ eV and $T_{B}=9998.3$ eV.}
	\label{SRCandNRCdiff}
\end{figure}
In Fig.~(\ref{SRCandNRCdiff}), we plot the two TDCSs (SRCBA and NRCBA) with respect to the symmetric coplanar geometry at relativistic energies. In the relativistic regime, by increasing the value of the incident kinetic energy ($10$ keV, $15$ keV, $20$ keV), we notice the shift of the maximum of the TDCS in
the SRCBA towards smaller values than $\theta_{B}=45^{\circ}$, as well as the fact that the SRCBA is always lower than the NRCBA.
\begin{figure}[h!]
\centering
	 \subfloat[]{\label{4TDCS250symdiff}\includegraphics[height=5cm,width=.4\linewidth]{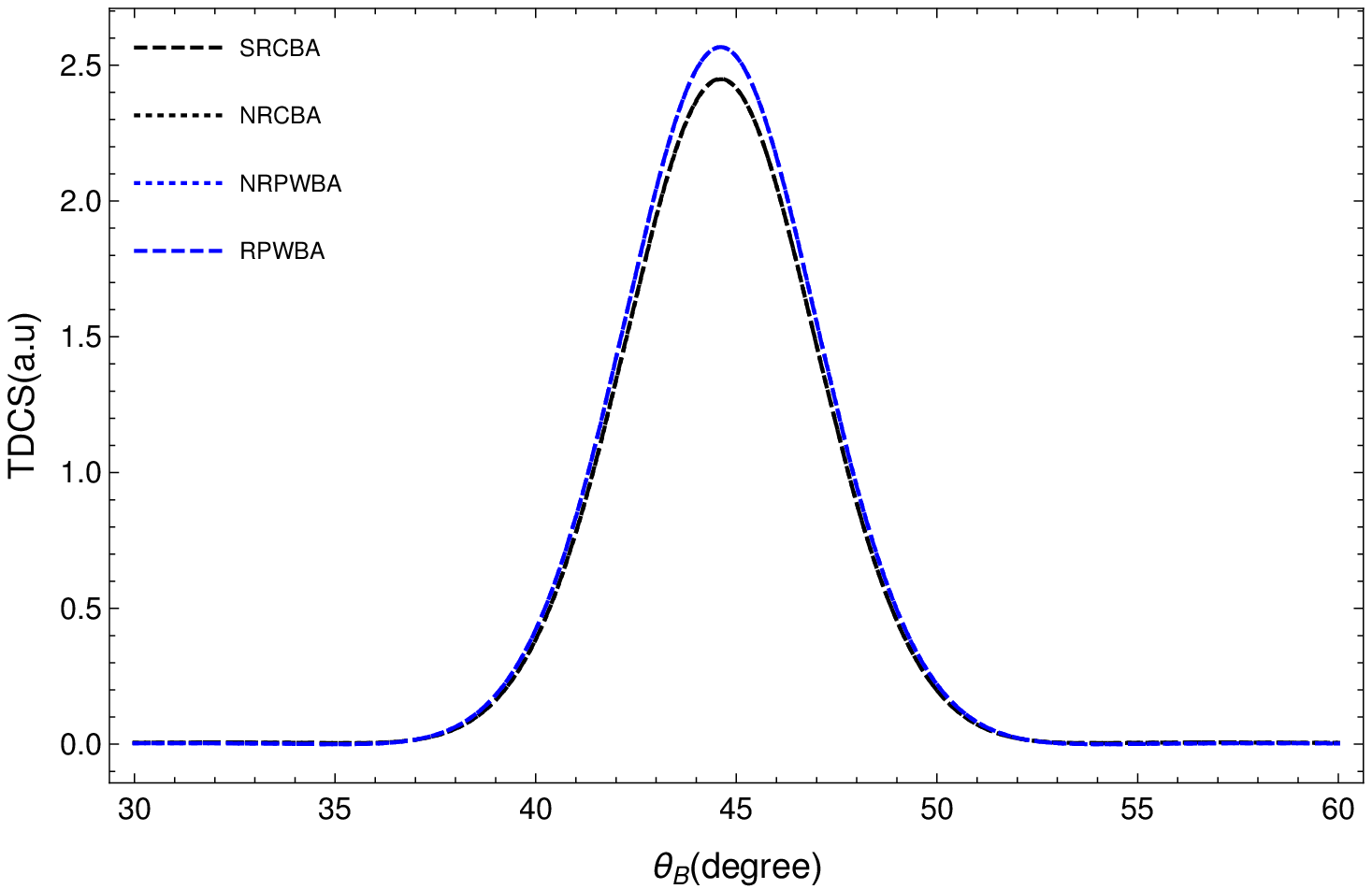}}\hspace*{.5cm}
	\subfloat[] {\label{4TDCS3250symconf}\includegraphics[height=5cm,width=.4\linewidth]{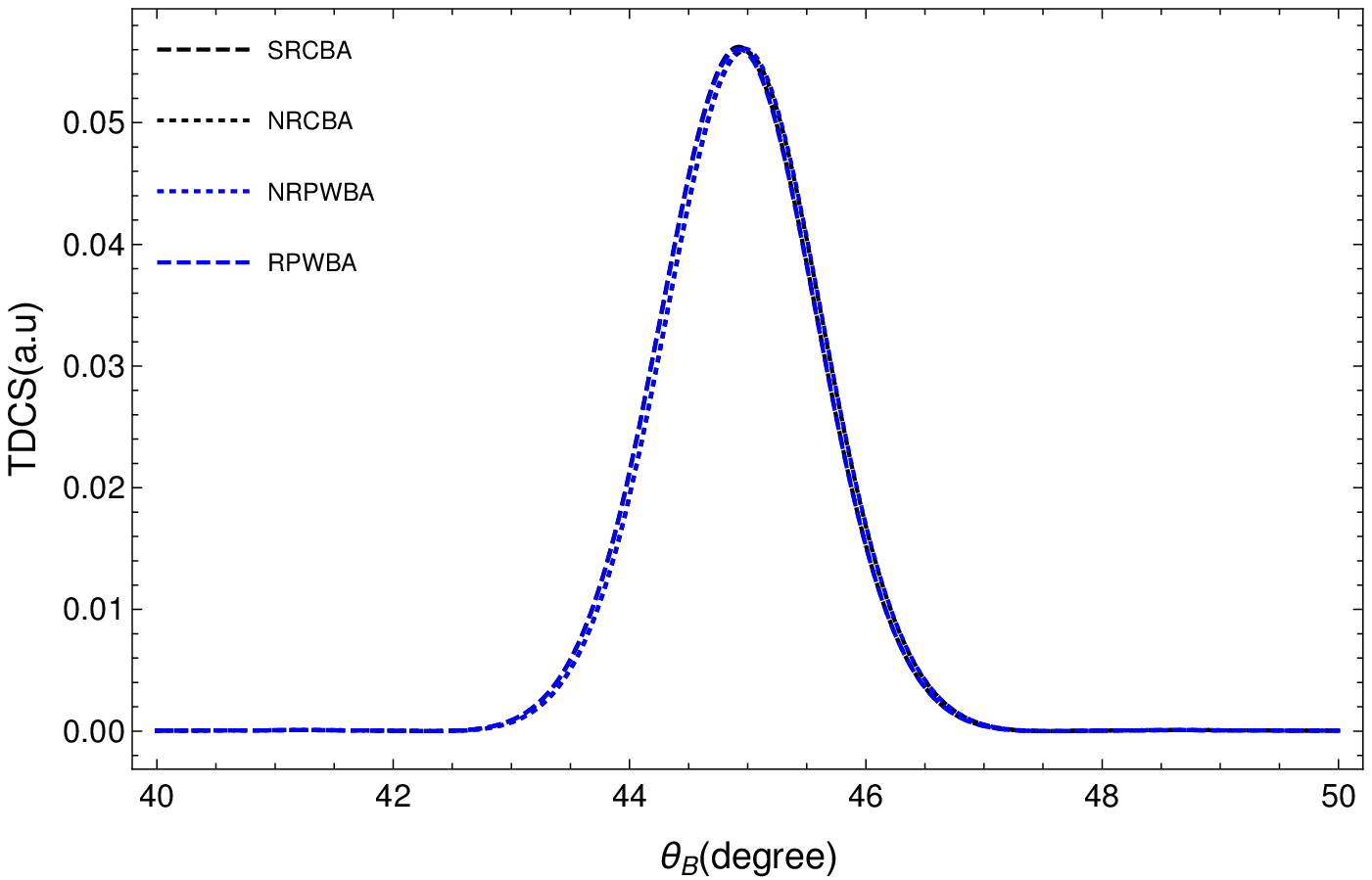}}
	\caption{The four TDCSs in symmetric coplanar geometry as a function of the ejection angle $\theta_{B}$ for the scattering angle $\theta_{f}=45^{\circ}$. The incident and the ejected electron kinetic energies are (a) $T_{i}=250$ eV and $T_{B}=123.3$ eV and (b) $T_{i}=3250$ eV and $T_{B}=1623.3$ eV.}
	\label{4TDCS}
\end{figure}
Recall that in the non-relativistic domain, we have already compared our results for the two asymmetric and symmetric coplanar geometries. In Fig.~(\ref{4TDCS}), we plot the RPWBA and the NRPWBA with the two TDCSs (SRCBA and NRCBA). We see that the NRCBA and the SRCBA are the same and they give a lower TDCS due to the fact that the ejected electron still feels Coulomb effect of the residual ion as much as its kinetic energy ($T_{B}=123.3$ eV in this case) is insufficient to cross the Coulomb barrier imposed by the residual ion. By increasing the kinetic energy of the incident and ejected electrons simultaneously (always checking the symmetric coplanar geometry), we find that the ejected electron begins slowly to escape from the Coulomb effect until it completely crosses it at kinetic energy $T_{B}=1623.3$ eV. In Fig.~(\ref{4TDCS3250symconf}), there is a very good
agreement between the four models and they produce the same results as the use of the Coulomb wave function is no longer necessary.
For the sake of comparison with the experimental results available in the literature for the total cross section, we attempted to calculate our total cross section in the RPWBA by performing the numerical integration of the TDCS (\ref{RPWBA}) over all outgoing scattering angles and energies. After that, the total cross section depends only on $E_{i}$, the incident electron kinetic energy.
\begin{equation}\label{tcs}
\begin{split}
\bar{\sigma}_{tot}^{RPWBA}&=4 \pi^{2} \int_{0}^{E_{B}^{max}/2} \frac{d\bar{\sigma}^{(RPWBA)}}{dE_{B}d\Omega_{B}d\Omega_{f}}dE_{B}\int_{0}^{\pi} \sin(\theta_{B})d\theta_{B}\int_{0}^{\pi} \sin(\theta_{f})d\theta_{f},
\end{split}
\end{equation}
where $E_{B}^{max}$ is the maximum value of the ejected electron kinetic energy. It is chosen, according to the kinetic energy conservation, to be $E_{B}^{max}=E_{i}+\mathcal{E}_{b}(2S)$ where $\mathcal{E}_{b}(2S)$ is the binding energy of the metastable 2S-state of atomic hydrogen given in (\ref{binding}). The division by $2$ in the maximum limit of integral over $E_{B}$ is inspired by the author Prasad \cite{prasad}. In Table \ref{tab1}, we compare our results (\ref{tcs}) with those obtained theoretically by Mukherjee \textit{et al.} \cite{mukherjee} using a rigorous distorted wave method in which the effects of both the initial and final channel distortions are taken into account. Table \ref{tab2} summarizes the comparison of our results with the experimental ones measured by Defrance \textit{et al.} using a crossed beam method where an electron beam intersects a beam of atomic hydrogen at $90^{\circ}$ \cite{defrance1981}.  The notable differences that appear between our results and Mukherjee's results or the experimental ones may be mainly due to two main factors, the first of which is that the experimental results are obtained by colliding a beam of electrons with a beam of hydrogen atoms, and thus extracting the total cross section data in terms of the center of mass energy, while we have assumed the hydrogen atom to be stable and bombarded it with an electron. This means that we are inputting the kinetic energy of the incoming electron. This difference in the input parameters will inevitably lead to different final results. The second is that the maximum limit of the ejected electron kinetic energy $E_{B}^{max}$ cannot be determined in an agreed and controlled method. Each one and the method he thinks is logical and follows to define this maximum value. This difference, along with the difference in the theoretical approach applied to study the ionization process, may constitute an obstacle to comparing the theoretical results with each other. Nevertheless, all results will be acceptable as long as they are approximately equal, at least, in order of magnitude about $10^{-16}\text{cm}^{2}$.
\begin{table}[ht]
\centering
\caption{Comparison of our results (\ref{tcs}) with the theoretical ones of Mukherjee \textit{et al.} \cite{mukherjee} for total cross section $\text{e}^{-}-\text{H(2S)}$ ionization.}
\begin{tabular}[t]{|l|*{11}{c|}}
\hline \multirow{2}{*}{Incident electron energy [eV]}&
\multicolumn{2}{c|}{ Total cross section $[10^{-16}\text{cm}^{2}]$}\\
\cline{2-3} &Theoretical results&Our results  \\
\hline
~~~~~~~~~~~~~~~5.1    &  4.665  &  2.43  \\
~~~~~~~~~~~~~~~7.65   &  8.416  &  7.13  \\
~~~~~~~~~~~~~~~10.2   &  10.03  &  9.48 \\
~~~~~~~~~~~~~~~13.6   &  9.424  &  10.36  \\
~~~~~~~~~~~~~~~17     &  8.672  &  10.19  \\
~~~~~~~~~~~~~~~20.4   &  7.968  &  9.67  \\
~~~~~~~~~~~~~~~30.6   &  6.608  &  7.93  \\
~~~~~~~~~~~~~~~40.8   &  5.360  &  6.56  \\
~~~~~~~~~~~~~~~68     &  3.792  &  4.32 \\
\hline
\end{tabular}
\label{tab1}
\end{table}
\begin{table}[ht]
\centering
\caption{Comparison of our results (\ref{tcs}) with the experimental ones of Defrance \textit{et al.} \cite{defrance1981} for total cross section $\text{e}^{-}-\text{H(2S)}$ ionization.}
\begin{tabular}[t]{|l|*{11}{c|}}
\hline \multirow{2}{*}{\shortstack{Center of mass \\energy [eV]}}&
\multicolumn{1}{c|}{ $\sigma_{tot}^{2S}$ $[10^{-16}\text{cm}^{2}]$}&\multirow{2}{*}{\shortstack{Incident electron \\energy [eV]}}&
\multicolumn{1}{c|}{$\bar{\sigma}_{tot}^{RPWBA}$ $[10^{-16}\text{cm}^{2}]$}\\
\cline{2-2} \cline{4-4}&Experimental results&&Our results  \\
\hline
~~~~~~~6.3    &  5.94  &  6.3   &  4.31  \\
~~~~~~~8.3    &  8.75  &  8.3   &  6.98  \\
~~~~~~~10.3   &  10.5  &  10.3  &  8.39  \\
~~~~~~~12.3   &  7.67  &  12.3  &  8.99  \\
~~~~~~~14.3   &  8.06  &  14.3  &  9.13  \\
~~~~~~~18.3   &  7.56  &  18.3  &  8.81  \\
~~~~~~~23.3   &  6.22  &  23.3  &  8.06  \\
~~~~~~~25.3   &  6.92  &  25.3  &  7.75  \\
~~~~~~~31.8   &  6.63  &  31.8  &  6.82  \\
\hline
\end{tabular}
\label{tab2}
\end{table}
\section{Conclusion}
In this work, we have calculated the triple differential cross sections (TDCS) for the ionization of hydrogen atom by electron impact in the metastable 2S-state for asymmetric and symmetric coplanar geometries. In the asymmetric coplanar geometry, we have compared our nonrelativistic results with those of other theories and found that the present model is very close to that obtained by Coulomb wave function at the scattering angles $\theta_{f}=3^{\circ}$ and $\theta_{f}=5^{\circ}$. In the symmetric coplanar geometry, a new nonrelativistic limit value is determined theoretically to be $4250$ eV, which is very different from that known for the ground state ($2700$ eV) \cite{taj}. Relativistic triple differential cross section have been evaluated within the relativistic model (RPWBA) in the first Born approximation. The consistency of this theoretical model is checked by taking the nonrelativistic limit. Semirelativistic TDCS in the SRPWBA gives nearly the same results, regardless of the kinetic energy of the incoming electron, as the RPWBA if the condition $Z\alpha\ll 1$ is satisfied. It is shown that the nonrelativistic formalism is no longer valid, in both geometries, for incident kinetic energies higher than $10$ keV, due to the spin and relativistic effects which begin to appear at high energies. Comparing our results for the two asymmetric and symmetric coplanar geometries, we found that the use of the Coulomb wave function to describe the ejected electron is no longer necessary as long as its kinetic energy $T_{B}\geq 1623.3$ eV. The validation of this work requires an experimental study.  We hope that our results should serve as a motivation to perform such collisions experiments in the future.


\begin{thebibliography}{99}
	\bibitem{ghoshdeb} S. Ghosh Deb, A. Biswas, and C. Sinha, J. Phys. B: At. Mol. Opt. Phys. \textbf{44}, 215201-1 (2011).\\
	\bibitem{taj} Y. Attaourti, S. Taj, and B. Manaut, Phys. Rev. A \textbf{71}, 062705 (2005).
	\bibitem{attaourti} Y. Attaourti and S. Taj, Phys. Rev. A \textbf{69}, 063411 (2004).
	\bibitem{nakel} W. Nakel and C. T. Whelan, Phys. Rep. \textbf{315}, 409 (1999).
	\bibitem{brauner1} M. Brauner, J. S. Briggs, and H. Klar, J. Phys. B: At. Mol. Opt. Phys. \textbf{19}, L325 (1986).
	\bibitem{brauner2} M. Brauner and J. S. Briggs, J. Phys. B: At. Mol. Opt. Phys. \textbf{24}, 2227 (1991).
	\bibitem{berkader} J. Berakder and H. Klar, J. Phys. B: At. Mol. Opt. Phys. \textbf{26}, 3891 (1993).
	\bibitem{kover1} A. Kover and G. Laricchia, Phys. Rev. Lett. \textbf{80}, 5309 (1998).
	\bibitem{kover2} C. Arcidiacono, A. Kover, and G. Laricchia, Phys. Rev. Lett. \textbf{95}, 223202 (2005).
	\bibitem{dorr} M. Dorr \textit{et al}, Phys. Rev. A \textbf{77}, 032717 (2008).
	\bibitem{dorn} A. Dorn, M. Dorr, B. Najjari, N. Haag, C. Dimopoulou, D. Nandi, and J. Ullrich, J. Electron Spectrosc. \textbf{161}, 2 (2007).
	\bibitem{ren} X. Ren, A. Dorn, and J. Ullrich, Phys. Rev. Lett. \textbf{101}, 093201 (2008).
	\bibitem{dixon1975} A. J. Dixon, A. Von Engel, and M. F. A. Harrison, Proc. R. Soc. Lond. A. \textbf{343}, 333 (1975).
	\bibitem{defrance1981} P. Defrance, W. Clays, A. Cornet, and G. Poulaert,  J. Phys. B: At. Mol. Phys. \textbf{14}, 111 (1981).
	\bibitem{hafid} H. Hafid, B. Joulakian, and C. Dal Cappello, J. Phys. B \textbf{26}, 3415 (1993).
	\bibitem{vucic} S. Vucic, R. M. Potvliege, and C. J. Joachain, Phys. Rev. A \textbf{35}, 1446 (1987).
	\bibitem{BBK} M. Brauner, J. S. Briggs, and H. Klar, J. Phys. B: At. Mol. Opt. Phys. \textbf{22}, 2265 (1989).
	\bibitem{dhar} S. Dhar, Aust. J. Phys. \textbf{49}, 937 (1996).
	\bibitem{das} J. N. Das and S. Dhar, Pramana J. Phys. \textbf{47}, 263 (1996).
	\bibitem{biswas} R. Biswas and C. Sinha, Nuovo Cimento D \textbf{16}, 571 (1994).
	\bibitem{ray} H. Ray and A. C. Roy, J. Phys. B: At. Mol. Opt. Phys. \textbf{21}, 3243 (1988).
	\bibitem{rmatrix} P. Descouvemont and D. Baye, Rep. Prog. Phys. \textbf{73}, 036301 (2010).
	\bibitem{ccc} I. Bray and D. V. Fursa, Phys. Rev. Lett. \textbf{76}, 2674 (1996).
	\bibitem{dwba} D. H. Madison and O. Al-Hagan, J. At. Mol. Opt. Phys. \textbf{2010}, 1 (2010).
	\bibitem{ehrhardt1969} H. Ehrhardt, M. Schulz, T. Tekaat, and K. Willmann, Phys. Rev. Lett. \textbf{22}, 89 (1969).
	\bibitem{amaldi1969} U. Amaldi, A. Egidi, R. Marconero, and G. Pizzella, Rev. Sci. Instrum. \textbf{40}, 1001 (1969).
	\bibitem{book1} J. Eichler and W. E. Meyerhof, \textit{Relativistic Atomic Collisions} (Academic, New York, 1995).
	\bibitem{integral} H. S. W. Massey and C. B. O. Mohr, Proc. R. Soc. London \textbf{140}, 613 (1933).
	\bibitem{greiner} W. Greiner and J. Reinhardt, \textit{Quantum Electrodynamics} (Springer-Verlag, Berlin, 1992).
	\bibitem{joachain} C. J. Joachain, \textit{Quantum Collision Theory}, (Edition North-Holland Publishing Company Amsterdam, 1975).	
	\bibitem{prasad} S. S. Prasad, Proc. Phys. Soc. \textbf{87}, 393 (1966).
	\bibitem{mukherjee} K. K. Mukherjee, K. B. Choudhury, N. R. Singh, P. S. Mazumdar and S. Brajamani, Aust. J. Phys. \textbf{42}, 475 (1989). 
\end{thebibliography}
\end{document}